\documentclass[twocolumn]{aastex631}

\usepackage{graphicx}
\usepackage{booktabs}

\DeclareRobustCommand
  \compactcdots{\mathinner{\cdotp\mkern-2mu\cdotp\mkern-2mu\cdotp}}

\newcounter{rtask}
\newcommand{\rtask}[1]{\refstepcounter{rtask}\label{#1}}

\begin{document}

\title{\large Evidence of a Disk-wind Origin for Fluorescent H\textsubscript{2} in Classical T Tauri Stars}

\author[0000-0003-2690-6241]{Matt Kalscheur}
\affiliation{Laboratory for Atmospheric and Space Physics, University of Colorado Boulder, Boulder, CO 80303, USA}

\author[0000-0002-1002-3674]{Kevin France}
\affiliation{Laboratory for Atmospheric and Space Physics, University of Colorado Boulder, Boulder, CO 80303, USA}

\author[0000-0002-9190-0113]{Brunella Nisini}
\affiliation{INAF - Osservatorio Astronomico di Roma, Via Frascati 33, 00078 Monte Porzio Catone, Italy}

\author[0000-0002-5094-2245]{P. Christian Schneider}
\affiliation{Hamburger Sternwarte, Gojenbergsweg 112, 21029 Hamburg, Germany}

\author[0000-0001-6410-2899]{Richard Alexander}
\affiliation{School of Physics \& Astronomy, University of Leicester, University Road, Leicester LE1 7RH, United Kingdom}

\author[0000-0001-6496-0252]{Jochen Eislöffel}
\affiliation{Thüringer Landessternwarte, Sternwarte 5, 07778 Tautenburg, Germany}

\author[0000-0002-3913-3746]{Justyn Campbell-White}
\affiliation{European Southern Observatory, Karl-Schwarzschild-Strasse 2, 85748 Garching bei M\"unchen, Germany}

\author[0000-0001-8385-9838]{Hsien Shang}
\affiliation{Institute of Astronomy and Astrophysics, Academia Sinica, Taipei 106216, Taiwan}

\author[0000-0002-8364-7795]{Manuele Gangi}
\affiliation{INAF - Osservatorio Astronomico di Roma, Via Frascati 33, 00078 Monte Porzio Catone, Italy}
\affiliation{ASI, Italian Space Agency, Via del Politecnico snc, 00133 Rome, Italy}

\author[0000-0003-0292-4832]{Zhen Guo}
\affiliation{Instituto de Física y Astronomía, Universidad de Valparaíso, Gran Bretaña 1111, Valparaíso, Chile}
\affiliation{Millennium Institute of Astrophysics, Nuncio Monseñor Sótero Sanz 100, Providencia, Santiago, Chile}

\author[0000-0002-0112-5900]{Seok-Jun Chang}
\affiliation{Max Planck Institute for Astrophysics, Karl-Schwarzschild-Strasse 1, 85748 Garching bei M\"unchen, Germany}

\correspondingauthor{Matt Kalscheur}
\email{Matt.Kalscheur@colorado.edu}

\begin{abstract}

We use FUV spectra of 36 T Tauri stars, predominately from \textit{Hubble Space Telescope}'s ULLYSES program, to examine the kinematic properties of fluorescent H\textsubscript{2} emission lines for evidence of disk outflows.  Leveraging improvements to the \textit{HST}-COS wavelength solution, we co-add isolated lines within four fluorescent progressions ([\textit{v'},\textit{J'}] = [1,4], [1,7], [0,2], and [3,16]) to improve signal-to-noise, and we fit each co-added line profile with one or two Gaussian components.  Of the high S/N line profiles (S/N $\geq$ 12 at the peak of the profile), over half are best fit with a combination of a broad and narrow Gaussian component.  For profiles of the [1,4] and [1,7] progressions, we find a systematic blue-shift of a few km s\textsuperscript{-1} between the broad and narrow centroid velocities and stellar radial velocities.  For the [0,2] progression, we find centroid velocities consistently blueshifted with respect to stellar radial velocities on the order of -5 km s\textsuperscript{-1} for the single and narrow components, and -10 km s\textsuperscript{-1} for the broad components.  Overall, the blueshifts observed in our sample suggest that the molecular gas traces an outflow from a disk wind in some sources, and not solely disk gas in Keplerian rotation.  The low-velocity systematic blue-shifts, and emitting radii as inferred from line FWHMs, observed in our sample are similar to those observed with optical [O I] surveys of T Tauri stars.  We estimate H\textsubscript{2} mass-loss rates of 10\textsuperscript{-9} to 10\textsuperscript{-11} $M_{\odot}$ yr\textsuperscript{-1}, but incomplete knowledge of wind parameters limits comparisons to global models.


\end{abstract}

\section{Introduction}

Circumstellar disks of gas and dust around pre-main-sequence stars typically disperse on the order of a few Myr in an inside-out fashion (e.g., \citealt{Fedele_2010, Ribas_2014}).  Mechanisms to explain this dispersal, and thus the evolution of the disk, include accretion onto the star, outflows in the form of jets and winds, and planet formation (e.g., \citealt{Armitage_2011, Alexander_2014}).

In order for disk material to accrete onto the star, angular momentum must be transported away from the inner disk.  The classic picture to explain this transport invokes turbulence driven by the magnetorotational instability (MRI; \citealt{Balbus_1991}).  However, this picture has been increasingly challenged by disk simulations which show that MRI-induced turbulence is suppressed in the planet-forming region of the disk (interior to 30 au; e.g., \citealt{Gammie_1996, Turner_2014}), and by ALMA observations which show that turbulence at tens of au is insufficient to drive accretion (e.g., \citealt{Teague_2016, Flaherty_2017}).

Instead, magnetohydrodynamic (MHD) disk winds, launched along magnetic field lines from the surface of the disk (see, e.g., \citealt{Blandford_1982, Pudritz_1986} for early theoretical work), are emerging as the preferred alternative to extract angular momentum from the disk (e.g., \citealt{Bai_2013, Tabone_2022}).  Local box (e.g., \citealt{Bai_2013}) and global (e.g., \citealt{Gressel_2015, Bethune_2017}) MHD simulations show that these winds arise naturally in the disk, can extend throughout the planet-forming region, and are sufficient to drive accretion at observed rates.  In addition to MHD winds, thermal photoevaporative (PE) winds are launched when high-energy photons from the star impart enough energy to unbind material from the disk surface (see, e.g., \citealt{Ercolano_2017} for a review).  They do not play a role in angular momentum transport, but are cited as a major mechanism in disk clearing after accretion has subsided in the later stages of disk evolution (e.g., \citealt{Alexander_2014, Pascucci_2023}).  As planets form out of the same material shaped by disk winds, an understanding of disk winds is important to informing theories of planet formation and evolution.

Spectroscopic tracers of disk winds in classical T Tauri stars (CTTSs) include atomic or weakly-ionized forbidden emission lines, mainly of oxygen, sulfur and neon in the optical and infrared spectral ranges (e.g., \citealt{Fang_2018, Banzatti_2019, Pascucci_2020, Nisini_2024}).  The outflowing emission is often blueshifted toward the observer, implying that part of the receding outflow is hidden by the circumstellar disk (e.g., \citealt{Jankovics_1983, Edwards_1987}).

Of the forbidden lines, the [O I] line at 6300 {\AA} is typically the brightest and most often studied.  It frequently exhibits both a high-velocity component (HVC) blueshifted to a few hundred km s\textsuperscript{-1} and a low-velocity component (LVC) blueshifted to a few km s\textsuperscript{-1} (e.g., \citealt{Kwan_1988, Hartigan_1995, Hirth_1997}).  These have been associated with magnetically-collimated jets of outflowing gas launched at small opening angles ($<$5\textdegree) near the star-disk interaction region ($\sim$0.1 au) (e.g., \citealt{Kwan_1988, Dougados_2000, Pascucci_2023}), and wide-angle ($\sim$30\textdegree) flows opening from near the disk surface (i.e., disk winds) (e.g., \citealt{Hartigan_1995, Natta_2014, Pascucci_2023}), respectively.  At high resolution ($\Delta v$ $\lesssim$ 5 km s\textsuperscript{-1}), the LVC can often be decomposed into a narrow Gaussian component (NC) accounting for the line peak, and a broad Gaussian component (BC) accounting for the broad line wings (e.g., \citealt{Rigliaco_2013, Simon_2016, Banzatti_2019}).  For the BC, inferred launching radii interior to 0.5 au, and correlations with the HVC and accretion, suggest an origin in a MHD wind (e.g., \citealt{Simon_2016, Banzatti_2019}).  As the NC is launched beyond 1 au, a MHD or photoevaporative wind origin is less clear, but correlations with the BC kinematics may also favor a MHD wind (\citealt{Banzatti_2019}).  In more evolved disks, NC peak velocities are often compatible with stellar radial velocities (RV), indicative of gas bound to the disk and not in a wind (e.g., \citealt{Nisini_2024}).  LVCs best fit by a single Gaussian component (SC) show evidence for the co-evolution of the line profile and the dispersion of the disk, with the [O I] emitting region receding to larger radii as the inner disk becomes optically thin (e.g., \citealt{McGinnis_2018, Banzatti_2019}).  HVCs and LVCs are also detected in [Ne II] 12.81 $\mu$m lines (\citealt{Pascucci_2020}), and increasingly in JWST spectro-imaging of [Ne II] and [Ne III] (e.g., \citealt{Bajaj_2024, Arulanantham_2024, Schwarz_2024}).

Molecular tracers of disk winds in CTTSs include the 1-0 S(1) line of H\textsubscript{2} at 2.12 $\mu$m (e.g., \citealt{Agra-Amboage_2014, Gangi_2020}), mid-infrared lines of H\textsubscript{2} in recent JWST spectro-imaging (e.g., \citealt{Arulanantham_2024, Schwarz_2024, Pascucci_2024}), and ro-vibrational transitions of CO near 4.7 $\mu$m (e.g., \citealt{Pontoppidan_2011, Brown_2013, Banzatti_2022}).  When detected, the 1-0 S(1) line of H\textsubscript{2} has been shown to be kinematically linked to the NC of the [O I] line at 6300 {\AA}, with clear blueshifted peaks in some disks (\citealt{Gangi_2020}).  This indicates an origin in a wind.  However, it is not detected alongside [O I] interior to 1 au.  The ro-vibrational transitions of CO are detected from emitting regions inside 1 au, with two-component (NC and BC) line profiles and systematic blue-shifts of a few km s\textsuperscript{-1} observed in some T Tauri disks (e.g., \citealt{Brown_2013, Banzatti_2022}).  More massive Herbig disks tend to produce double-peaked CO line profiles, indicative of gas bound to rings in Keplerian rotation, not gas in an outflow (\citealt{Banzatti_2022}).

Fluorescent H\textsubscript{2} emission in the far ultraviolet (FUV), pumped by photons spanning the width of the Lyman-alpha (Ly-$\alpha$) emission line (\citealt{Herczeg_2002}), is a tracer of the inner gas disk in CTTSs (e.g., \citealt{France_2012, France_2023}).  It can arise within 0.1 au of the parent star and is detected in virtually all pre-main-sequence systems surveyed to date (\citealt{France_2023}).  It shows correlations with the NC of [O I] emission at 6300 {\AA} (\citealt{Gangi_2023}), but most existing analyses do not consider a disk-wind origin (e.g., \citealt{Gangi_2023, France_2023}).  Fluorescent H\textsubscript{2} line profiles are fit by a single Gaussian component and their centroid velocities are taken to be coincident with stellar RVs, within the original 15 km s\textsuperscript{-1} wavelength accuracy of \textit{Hubble Space Telescope}'s Cosmic Origins Spectrograph (\textit{HST}-COS; \citealt{Oliveira_2010, Green_2012}).  However, fluorescent H\textsubscript{2} emission has been observed as an outflow component in FUV imaging of at least one CTTS (\citealt{Schneider_2013}).

In this paper, we revisit the use of fluorescent H\textsubscript{2} as a spectroscopic tracer of disk winds in a sample of 36 CTTSs.  This is made possible by improvement of the \textit{HST}-COS wavelength accuracy to $<$5 km s\textsuperscript{-1} (e.g., \citealt{Plesha_2019, Plesha_2022}), allowing for more accurate centroid velocities and the creation of high-S/N line profiles, and recent investment in \textit{HST}'s Ultraviolet Legacy Library of Young Stars as Essential Standards (ULLYSES) program.  To improve the \textit{HST}-COS wavelength accuracy, the dispersion coefficients and zero points were re-derived through cross-correlation of on-orbit COS spectra with more accurate ($\sim$2 km s\textsuperscript{-1}) \textit{HST}-STIS (Space Telescope Imaging Spectrograph; \citealt{Woodgate_1998}) echelle spectra, and fitting of the dispersion terms to ray-trace models (\citealt{Plesha_2018}).  The ULLYSES program obtained UV spectroscopic observations of 71 T Tauri stars, and produced new data products from archival UV observations of additional low-mass stars (see \citealt{Roman-Duval_2020} for an overview).  Through this new analysis, we better place H\textsubscript{2}, the primary mass component of gas-rich disks, into the framework of existing atomic and molecular tracers of disk outflows.

The paper is organized as follows: In Section \ref{sec:2}, we discuss the archival \textit{HST}-COS spectra, predominately from ULLYSES Data Release 6 (DR6), and initial target selection.  In Section \ref{sec:3}, we describe the analysis of the fluorescent H\textsubscript{2} line profiles, and measurement of the H\textsubscript{2} emitting radius.  We present the results of our kinematic analysis in Section \ref{sec:4}, including correlations with disk inclination and the infrared index.  In Section \ref{sec:5}, we compare the results to those from the 6300 {\AA} line of [O I], both in the ensemble and on the level of a few individual disks.  In Section \ref{sec:6}, we discuss the origin and evolution of fluorescent H\textsubscript{2} in the context of our analysis.  We also produce initial estimates of H\textsubscript{2} mass loss in those disks with clear indications of H\textsubscript{2} in a wind.  We conclude in Section \ref{sec:7}.

\section{Archival \textit{HST}-COS Spectra and Sample Selection} \label{sec:2}

ULLYSES DR6\footnote{\url{https://ullyses.stsci.edu/ullyses-dr6.html}} includes 108 T Tauri stars with UV spectra, 85 of which were observed with both the G130M and G160M grating modes of \textit{HST}-COS.  These modes provide a velocity resolution between 15 and 25 km s\textsuperscript{-1} (\citealt{Soderblom_2023}), and continuous spectral coverage in the FUV from approximately 1130 to 1790 {\AA}, varying slightly with the specific central wavelength setting employed.  Observations made as part of the ULLYSES program were obtained primarily during \textit{HST} Cycle 28 and 29 (2020 to 2022), with additional archival observations made as early as 2009 (e.g., \citealt{Brown_2010, France_2012}).  General details on the observing strategy and standard data reduction are available in the ODYSSEUS overview paper (\citealt{Espaillat_2022}).  ODYSSEUS is a community effort to uniformly and systematically analyze ULLYSES spectra to better understand the accretion, outflow and evolutionary properties of CTTSs.

To augment the initial list, we also include six bright T Tauri targets (AK Sco, BP Tau, GM Aur, SU Aur, TW Hya, and V4046 Sgr) with archival observations (PIDs 11533, 11616, 12036, 12315, and 13372; PIs: Green, Herczeg, Green, Guenther, and Gómez de Castro), but not included in ULLYSES DR6.

\begin{figure}
  \centering
  \includegraphics[width=0.87\linewidth]{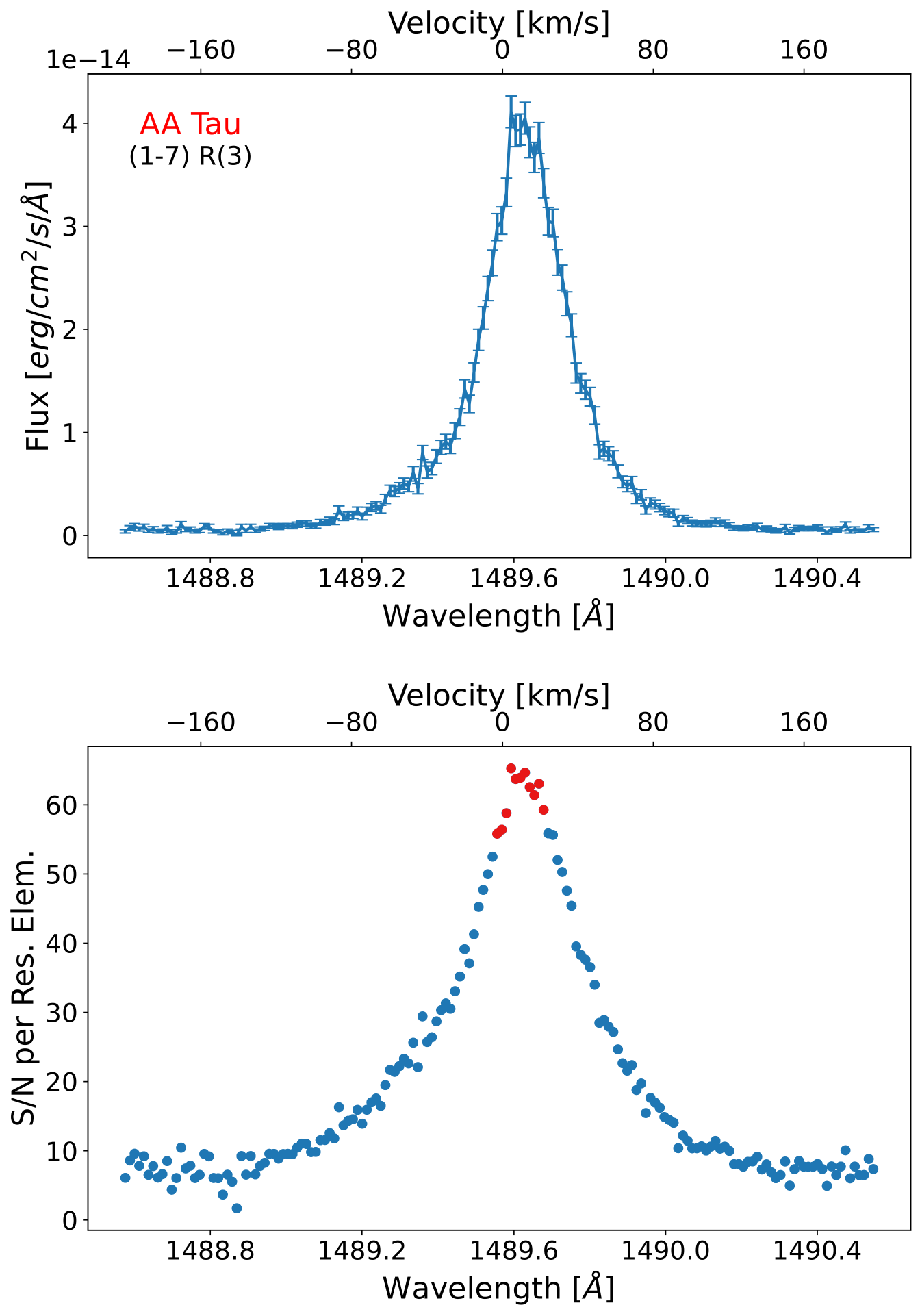}
  \caption{Example flux (top panel) and S/N (bottom panel) line profiles for the $B^{1}\Sigma^{+}_{u}-X^{1}\Sigma^{+}_{g}$ (1–7) R(3) fluorescent emission line of H\textsubscript{2} in ULLYSES target AA Tau.  Zero velocity is with respect to the rest wavelength of the line.  Average S/N per resolution element at the peak (denoted in red) of the profile is 61.25, the highest of the ULLYSES targets.}
  \label{fig:1}
\end{figure}

To produce a final sample for analysis, we make a cut on the 91 targets with both G130M and G160M observations, dropping those with average S/N $<$ 15 per resolution element\footnote{Resolution elements represent the actual spectral resolution of the data.  They are distinguished from the digitized pixels ($\sim$6 per resolution element) of COS's FUV detector (\citealt{Osterman_2011}).} at the peak of the bright and isolated $B^{1}\Sigma^{+}_{u}-X^{1}\Sigma^{+}_{g}$ (1–7) R(3) ($\lambda$\textsubscript{lab} = 1489.57 {\AA}) fluorescent emission line of H\textsubscript{2}.  The peak is taken to include the five brightest pixels on either side ($\pm$12 km s\textsuperscript{-1}) of observed line center.  Example flux and S/N profiles for the (1-7) R(3) line of AA Tau are shown in Figure \ref{fig:1}.

Of the remaining targets, we drop three (HD 104237, HD 142527 and HK Ori) which are more appropriately classified as Herbig Ae/Be stars (\citealt{Wenger_2000}), DR Tau, and RW Aur.  DR Tau exhibits atypical H\textsubscript{2} line profiles possibly generated in the circumstellar medium (e.g., \citealt{France_2014b}), while RW Aur has a redshifted, high-velocity H\textsubscript{2} outflow component rarely seen in other T Tauri stars (e.g., \citealt{France_2014a}).

\begin{table*}[t!]
\centering
\textbf{Table 1} \\
\text{Source Properties} \\
\smallskip
 \begin{tabular*}{\textwidth}{c @{\extracolsep{\fill}} cccccc}
 \hline
 \hline
 Name & $M$\textsubscript{\textsubscript{*}} & $\dot M$\textsubscript{acc} & $i$\textsubscript{disk} & $n$\textsubscript{13-31} & $v$\textsubscript{rad} & References \\
 & ($M_{\odot}$) & (10\textsuperscript{-8} $M_{\odot}$ yr\textsuperscript{-1}) & (deg) & & (km s\textsuperscript{-1}) & \\
 \hline
 AA Tau & 0.80 & 0.33 & 59 & -0.36 & 17.86 $\pm$ 0.01 & 7, 12, 27, 28, 29 \\
 AK Sco & 1.35 & 0.09 & 68 & 0.45 & -1.3 $\pm$ 2.6 & *, 15, 30, 31 \\
 BP Tau & 0.48 & 0.56 & 30 & -0.36 & 16.09 $\pm$ 0.16 & 7, 23, 24, 32 \\
 CS Cha & 1.05 & 1.20 & 37 & 2.89 & 15.23 $\pm$ 0.37 & 1, 2, 16 \\
 DE Tau & 0.26 & 1.02 & 34 & -0.13 & 14.67 $\pm$ 0.44 & 17, 23, 24, 33 \\
 DF Tau & 0.39 & 2.82 & 85 & -1.09 & 14.80 $\pm$ 0.54 & 7, 23, 24, 34 \\
 DM Tau & 0.50 & 0.29 & 35 & 1.29 & 18.61 $\pm$ 0.01 & 7, 9, 27, 35 \\
 ECHA J0843.3-7915 (ET Cha) & 0.20 & 0.08 & 60 & -0.20 & 22 & 2, 3, 36, 37 \\
 GM Aur & 1.19 & 0.51 & 55 & 1.75 & 15.98 $\pm$ 0.37 & 7, 23, 24, 38 \\
 HN Tau & 1.58 & 4.90 & 50 & -0.62 & 27.34 $\pm$ 2.25 & 7, 24, 25, 39 \\
 MY Lup & 1.06 & 1.02 & 72 & 0.18 & 4.4 $\pm$ 2.1 & *, 5, 26, 40, 41 \\
 RECX 11 (EP Cha) & 0.83 & 0.02 & 70 & -0.80 & 18 & 2, 4, 36, 37 \\
 RX J1556.1-3655 (WRAY 15-1384) & 0.47 & 1.48 & 53 & $\compactcdots$ & 2.6 $\pm$ 1.2 & 1, 5, 26 \\
 RX J1842.9-3532 (WRAY 15-1880) & 1.14 & 0.16 & 32 & 0.72 & -5.66 $\pm$ 2.21 & 6, 7, 8 \\
 RX J1852.3-3700 & 1.05 & 0.20 & 30 & 2.63 & -1.46 $\pm$ 2.21 & 6, 7, 8 \\
 RY Lup & 1.53 & 0.69 & 68 & 0.87 & 1.3 $\pm$ 2.0 & 1, 5, 7, 26 \\
 SSTc2d J160830.7-382827 & 1.53 & 0.05 & 72 & $\compactcdots$ & 1.2 $\pm$ 1.9 & 1, 5, 26 \\
 SSTc2d J161243.8-381503 & 0.46 & 0.08 & 43 & $\compactcdots$ & -2.3 $\pm$ 2.4 & 1, 5, 26 \\
 SU Aur & 2.22 & 1.74 & 62 & 0.74 & 14.26 $\pm$ 0.05 & 9, 17, 23, 42 \\
 SY Cha & 0.78 & 0.04 & 51 & -0.16 & 15.69 $\pm$ 0.27 & *, 18, 43, 44 \\
 Sz 45 & 0.56 & 0.51 & 43 & 0.98 & 14.92 $\pm$ 0.01 & *, 9, 43, 44 \\
 Sz 71 (GW Lup) & 0.42 & 0.09 & 40 & -0.44 & -3.3 $\pm$ 1.9 & *, 5, 26, 45 \\
 Sz 75 (GQ Lup) & 0.89 & 2.09 & 60 & -0.16 & -3.6 $\pm$ 1.3 & *, 5, 10, 26 \\
 Sz 82 (IM Lup) & 1.13 & 0.89 & 48 & -0.30 & -0.5 $\pm$ 1.3 & *, 5, 10, 26 \\
 Sz 98 (V1279 Sco) & 0.74 & 2.57 & 46 & -0.56 & -1.4 $\pm$ 2.1 & 5, 7, 26, 44 \\
 Sz 100 & 0.14 & 0.01 & 47 & -0.86 & 2.7 $\pm$ 2.5 & *, 5, 26, 44 \\
 Sz 102 & 0.24 & 0.08 & 53 & 0.64 & 21.6 $\pm$ 12.4 & 5, 7, 40, 44 \\
 Sz 103 & 0.24 & 0.05 & 50 & 0.17 & 1.4 $\pm$ 2.2 & *, 5, 26, 44 \\
 Sz 111 & 0.45 & 0.04 & 54 & $\compactcdots$ & -1.2 $\pm$ 2.1 & 5, 26, 44 \\
 Sz 114 (V908 Sco) & 0.22 & 0.06 & 6 & 0.01 & 4.0 $\pm$ 2.4 & *, 5, 26, 44 \\
 Sz 129 & 0.82 & 0.47 & 31 & $\compactcdots$ & 3.2 $\pm$ 2.5 & 5, 26, 44 \\
 TW Hya & 0.60 & 0.02 & 4 & 0.96 & 12.34 $\pm$ 0.00 & 7, 14, 19, 46 \\
 UX Tau A & 1.36 & 0.26 & 35 & 1.82 & 19.74 $\pm$ 0.35 & 6, 7, 23, 24, 47 \\
 V4046 Sgr (HD 319139) & 0.86 & 1.30 & 36 & 0.32 & -5.7 $\pm$ 0.2 & 2, 13, 21 \\
 V510 Ori & 0.76 & 0.58 & $\compactcdots$ & -0.18 & 33.31 $\pm$ 0.24 & *, 1, 22 \\
 VZ Cha & 0.80 & 6.61 & 25 & -0.91 & 11.0 $\pm$ 1.0 & 9, 11, 17, 20 \\
 \hline
 \end{tabular*}
\begin{flushleft}
\textbf{Notes.} Typical uncertainties for $M$\textsubscript{\textsubscript{*}}, $\dot M$\textsubscript{acc} and $i$\textsubscript{disk} are 0.15 dex, an order of magnitude and 10 degrees, respectively.  $n$\textsubscript{13-31} values for sources that include a * in their reference lists were calculated in this work.  References for the remaining star and disk parameter values: (1) \citet{France_2023}, (2) \citet{Arulanantham_2021}, (3) \citet{Woitke_2011}, (4) \citet{Ardila_2013}, (5) \citet{Frasca_2017}, (6) \citet{Francis_2020}, (7) \citet{Banzatti_2019}, (8) \citet{White_2007}, (9) \citet{Nguyen_2012}, (10) \citet{Long_2022}, (11) \citet{Pascucci_2020}, (12) \citet{Jonsson_2020}, (13) \citet{France_2017}, (14) \citet{France_2012}, (15) \citet{Gontcharov_2006}, (16) \citet{Hourihane_2023}, (17) \citet{Furlan_2009}, (18) \citet{Sacco_2017}, (19) \citet{Soubiran_2018}, (20) \citet{Banzatti_2017}, (21) \citet{Donati_2011}, (22) \citet{Tarricq_2021}, (23) \citet{Gangi_2022}, (24) \citet{Nisini_2024}, (25) \citet{Alcala_2021}, (26) \citet{Alcala_2019}, (27) \citet{Ingleby_2013}, (28) \citet{Gullbring_1998}, (29) \citet{Loomis_2017}, (30) \citet{Alencar_2003}, (31) \citet{Gomez_de_Castro_2009}, (32) \citet{Simon_2000}, (33) \citet{Simon_2019}, (34) \citet{Johns-Krull_2001}, (35) \citet{Kudo_2018}, (36) \citet{Rugel_2018}, (37) \citet{Lawson_2004}, (38) \citet{Andrews_2011}, (39) \citet{Simon_2017}, (40) \citet{Alcala_2017}, (41) \citet{Huang_2018}, (42) \citet{Akeson_2002}, (43) \citet{Manara_2017}, (44) \citet{Hendler_2020}, (45) \citet{Ansdell_2016}, (46) \citet{Pontoppidan_2008}, (47) \citet{Francis_2020}.
\end{flushleft}
\rtask{table:1}
\end{table*}

The final sample consists of 36 CTTS sources, with the relevant star and disk parameters reported in Table \ref{table:1}.  They come primarily from three star-forming regions (Chamaeleon I, Lupus and Taurus-Auriga) within 200 pc of the Solar System.  Stellar masses range between 0.14 and 2.22 $M_{\odot}$.  Outer disk inclinations, mostly from spatially-resolved ALMA images, vary between 4 (nearly face-on) and 85 (nearly edge-on) degrees.  The $n$\textsubscript{13-31} infrared index (as defined by, e.g., \citealt{Furlan_2009}) measures the slope of the spectral energy distribution between 31 and 13 $\mu$m.  It is a proxy for dust evolution in the inner disk, with values above $\sim$0 indicating dust depletion and the possible presence of an inner cavity (i.e., a transition disk).  For the 11 sources without literature values, we calculate $n$\textsubscript{13-31} indices from medium-resolution Spitzer spectra downloaded from the online CASSIS database (\citealt{Lebouteiller_2015}), utilizing the specific wavelength ranges defined in \citet{Banzatti_2019}.  Literature values of stellar RV are determined mainly through cross-correlation of optical photospheric lines with stellar models (e.g., \citealt{Banzatti_2019, Nisini_2024}).  The FUV flux measured by \textit{HST}-COS does not provide adequate stellar atmospheric lines to determine RVs self-consistently from the archival data.

\section{Data Reduction and Analysis} \label{sec:3}

We analyze bright and isolated (i.e., free of contamination from nearby lines) emission lines from four fluorescent progressions ([\textit{v'},\textit{J'}] = [1,4], [1,7], [0,2], and [3,16]) of H\textsubscript{2}.  A progression is a collection of individual emission lines pumped by Ly-$\alpha$ to common upper ro-vibrational states.  Relevant line parameters are summarized in Table \ref{table:2}.  The [1,4] and [1,7] progressions are pumped by Ly-$\alpha$ absorbing transitions within +100 km s\textsuperscript{-1} of Ly-$\alpha$ line center.  The [0,2] progression is pumped by Ly-$\alpha$'s red wing (+486 km s\textsuperscript{-1} from line center), and [3,16] by an absorbing transition to the blue (-296 km s\textsuperscript{-1} from line center).

\begin{table}
\centering
\textbf{Table 2} \\
\text{Selected H\textsubscript{2} Emission Lines} \\
\smallskip
 \begin{tabular*}{\linewidth}{c @{\extracolsep{\fill}} cccc}
 \hline
 \hline
 Line ID & $\lambda$\textsubscript{lab} & $B$\textsubscript{mn}\textsuperscript{(a)} & [\textit{v'},\textit{J'}]\textsuperscript{(b)} & $\lambda$\textsubscript{pump}\textsuperscript{(c)}  \\
 & (\AA) & & & (\AA) \\
 \hline
 (1–6) R(3) & 1431.01 & 0.058 & [1,4] & 1216.07 \\
 (1–6) P(5) & 1446.12 & 0.083 & [1,4] & 1216.07 \\
 (1–7) R(3) & 1489.57 & 0.094 & [1,4] & 1216.07 \\
 (1–7) P(5) & 1504.76 & 0.115 & [1,4] & 1216.07\medskip \\
 (1–6) R(6) & 1442.87 & 0.055 & [1,7] & 1215.73 \\
 (1–6) P(8) & 1467.08 & 0.080 & [1,7] & 1215.73 \\
 (1–7) R(6) & 1500.45 & 0.101 & [1,7] & 1215.73\medskip \\
 (0–4) P(3) & 1342.26 & 0.148 & [0,2] & 1217.64\medskip \\
 (3–5) R(15) & 1418.23 & 0.050 & [3,16] & 1214.47 \\
 (3–9) R(15) & 1593.26 & 0.122 & [3,16] & 1214.47 \\
 \hline
 \end{tabular*}
\begin{flushleft}
\textbf{Notes.} (a) Ratio of the line transition probability to the total transition probability out of state [\textit{v'},\textit{J'}].  (b) \textit{v'} and \textit{J'} refer to the vibrational and rotational quantum numbers in the excited ($B^{1}\Sigma_{u}^{+}$) electronic state of H\textsubscript{2}.  (c) Wavelength of the transition along the Ly-$\alpha$ profile that pumps the H\textsubscript{2}.
\end{flushleft}
\rtask{table:2}
\end{table}

\begin{figure}
  \centering
  \includegraphics[width=0.95\linewidth]{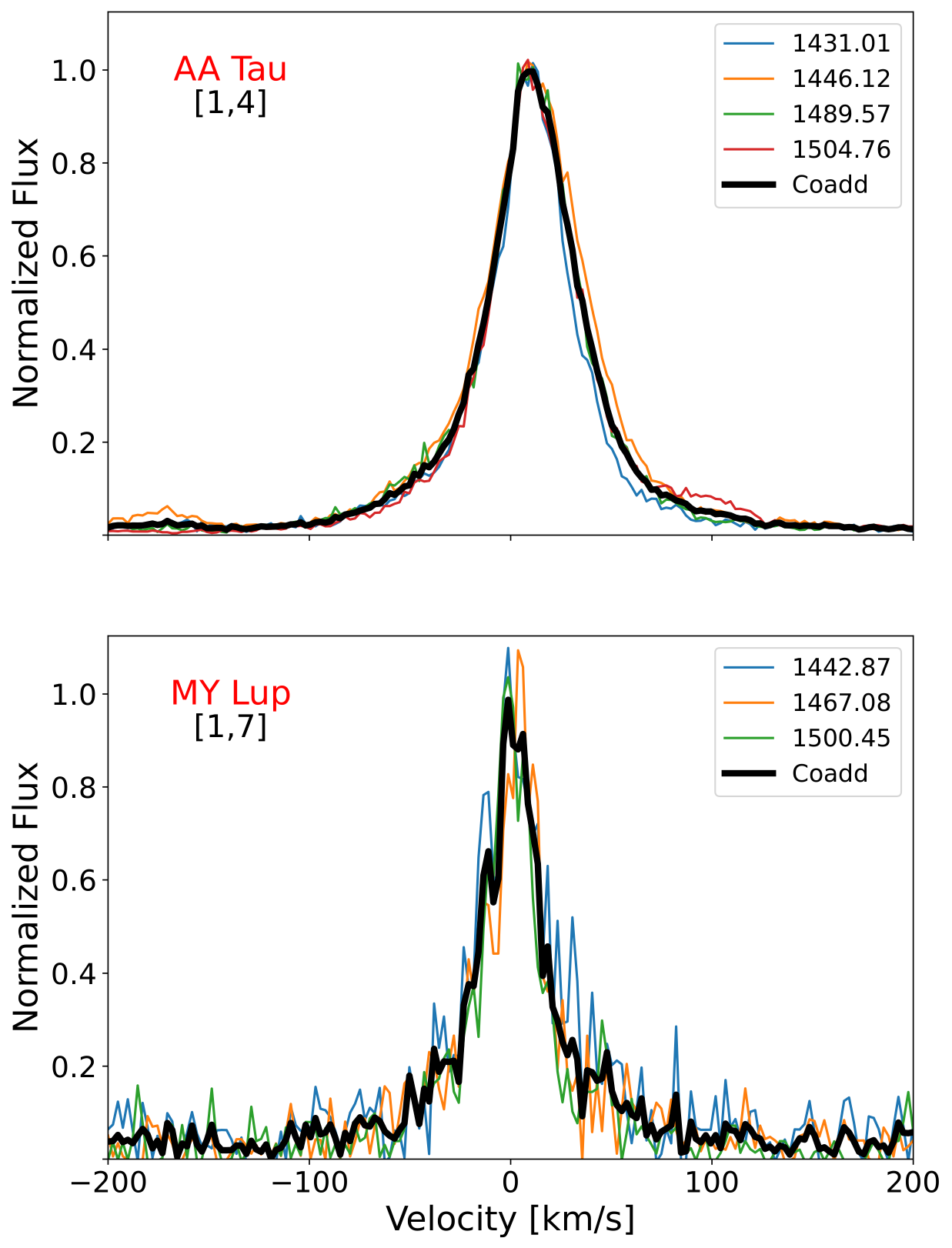}
  \caption{Example individual and co-added emission line profiles of the [1,4] progression for a high-S/N ($\sim$126 at the peak of the co-added profile) disk (top panel), and of the [1,7] progression for a low-S/N ($\sim$16) disk (bottom panel).  The averaged profile is indicated by the thick black line.  Zero velocity is with respect to the rest wavelength of line center.}
  \label{fig:2}
\end{figure}

To improve S/N, we co-add emission lines within each progression.  Before performing the co-addition, we convert line wavelengths into velocity space using the rest wavelengths of respective line centers.  To account for potential differences in the wavelength accuracy, we shift the peak of each line (taken to be the median of the 11 brightest pixels near line center) to match the peak of the brightest individual line (i.e., that line with the highest branching ratio) in each progression, rarely exceeding shifts of a single pixel ($\sim$2.3 km s\textsuperscript{-1}).  We normalize the flux of each line by the mean flux of its three brightest pixels.  The co-addition is thus the average of the individual normalized profiles of each progression, with each line contributing equally to the co-added line shape.  Example profiles representing ``high"- and ``low"-S/N cases are shown in Figure \ref{fig:2}.  We refer to the co-added lines as [1,4], [1,7] and [3,16], respectively.  As the [0,2] progression contains just a single bright line free of contamination in the majority of the disks (e.g., the bright (0–5)P(3) line of $\lambda$\textsubscript{lab} = 1402.65 {\AA} is often blended with a Si IV line of $\lambda$\textsubscript{lab} = 1402.77 {\AA}), we do not perform a co-addition.  We exclude co-added profiles with S/N $<$ 15 per resolution element at the peak of [1,7], and S/N $<$ 12 per resolution element at the peak of [0,2] and [3,16].  We are more permissive with the fainter [0,2] and [3,16] progressions so as not to exclude several profiles which appear amenable to model fitting.

We model each co-added line profile as a composite of either one or two Gaussian functions and a linear continuum, and convolve the model with the \textit{HST}-COS line-spread function (LSF).  The on-orbit shape of the LSF, introduced by polishing errors on the \textit{HST} primary and secondary mirrors, redistributes power from line center to non-Gaussian line wings, with more pronounced effect for narrow lines at short wavelengths (\citealt{Ghavamian_2009}).  As the shape of the LSF varies with wavelength, it is sampled every $\sim$5 {\AA} for each \textit{HST}-COS FUV grating and central wavelength setting, at every Lifetime Position (LP).\footnote{To mitigate the effect of gain sag, \textit{HST}-COS FUV spectra are recorded at different positions on the detector over time, known as Lifetime Positions (\citealt{Soderblom_2022}).}

For the [1,4], [1,7] and [3,16] progressions, we choose the G160M/1611 set of LSFs corresponding to LP4.  For the [0,2] progression, its sole emission line falls on the G130M grating, and we choose the G130M/1291 set of LSFs also corresponding to LP4.  The central wavelength settings (1611 and 1291) and LP that we select represent the instrument modes most frequently utilized in the ULLYSES observations.  To perform the convolution, we average the three individual LSF kernels closest to the mean of the rest wavelengths pertaining to the individual emission lines of each progression (e.g., average of the three LSFs sampled nearest to 1468 {\AA} for the [1,4] progression).\footnote{Parameter measurements from the fitting routine are relatively insensitive to the specific LSF kernel(s) chosen.  For example, convolving the model with the individual LSF kernel closest to the most blue (1428 {\AA}) or most red (1508 {\AA}) emission line of [1,4] results in FWHM measurements that vary less than 1\% for a typical ($\sim$50 km s\textsuperscript{-1}) Gaussian component width.}

We fit the composite model, convolved with the LSF, to the data using the Python package \textit{emcee}.\footnote{\textit{emcee} is an implementation of an affine-invariant ensemble Markov chain Monte Carlo (MCMC) sampler (\citealt{Foreman-Mackey_2013}).  In general, MCMC is used as a maximum likelihood estimator wherein parameter values of model components are iteratively compared to the data until convergence is reached.}  We initialize 64 walkers (essentially separate Markov chains) with generic guesses of the amplitude, location (i.e., the centroid velocity) and FWHM of the Gaussian component(s), and the slope and intercept of the linear continuum.  We iterate each walker through a 2048 step burn-in period (to reduce bias associated with the initial parameter estimates) and a 2048 step production (or fitting) period.  The outcome of the fitting period provides measurements and uncertainties for each parameter (either 5 or 8 in total) of the model components.  We take the measurement values from the median value of the samples in each parameter's marginalized distribution, and 1$\sigma$ uncertainties from the 16 and 84th percentile values.

We perform separate one- and two-component Gaussian fits for each co-added profile, keeping the model whose median fit best minimizes the Bayesian information criterion (BIC) (see \citealt{Arulanantham_2018} for this method applied to fluorescent H\textsubscript{2} emission in RY Lupi) and keeps the reduced $\chi^{2}$ closest to 1.  For those profiles with similar BIC ($\Delta$BIC $<$ 10) and reduced $\chi^{2}$ values for both models (approximately 1/4 of the sample for each progression), we choose a model based primarily on visual inspection of the model components (e.g., Are the Gaussian components of a two-component model distinct?  Is the amplitude of the broad Gaussian component at least 5\% of the model peak, and does it avoid an incidental fit to a continuum feature?).

\begin{figure}
  \centering
  \includegraphics[width=0.95\linewidth]{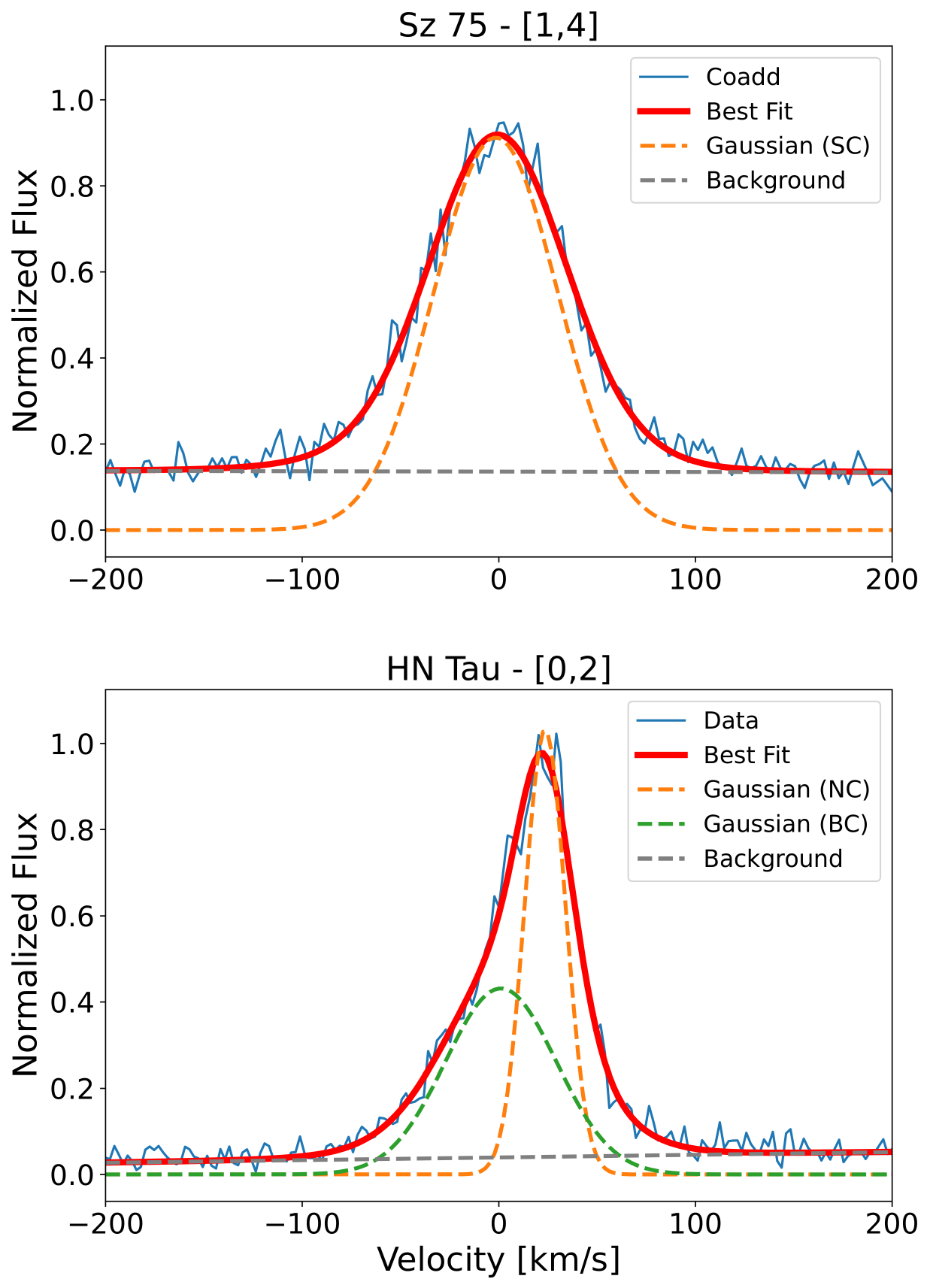}
  \caption{Representative examples of one- (top panel) and two-component (bottom panel) model fits to co-added fluoresecent H\textsubscript{2} line profiles.  The co-added profile is in blue, best-fit model in red, and linear continuum fit in gray dashes.  The sum of the broad (green dashes) and narrow (orange dashes) Gaussian components exceeds the height of the best-fit model because the best-fit model is convolved with the \textit{HST}-COS LSF (not shown).  Velocities are not corrected for stellar RVs.}
  \label{fig:3}
\end{figure}

For profiles best fit by a single Gaussian component and linear continuum, we refer to the single Gaussian component as SC.  For profiles best fit by two Gaussian components and a linear continuum, we refer to the narrower Gaussian component as NC, and the broader Gaussian component as BC.  This follows the procedure described in \citet{Banzatti_2019} for [O I] 6300 {\AA}, wherein Gaussian model components are labelled with respect to the number of components in their respective best-fit models, and not solely according to their specific FWHMs (as in, e.g., \citealt{Simon_2016}).  As the fluorescent H\textsubscript{2} line profiles do not exhibit evidence of jets, we do not require additional labels.  Our classificaton scheme is illustrated in Figure \ref{fig:3}.

Finally, we infer average H\textsubscript{2} emitting radii under the assumption that the width of the Gaussian components is dominated by kinematic broadening, expressed as

\begin{equation}
    R(\textrm{H\textsubscript{2}}) = G M_{\odot} \left(\frac{2\sin{(i)}}{\textrm{FWHM}}\right)^{2},
\end{equation}

\vspace{7.5pt}

\noindent where $M_{\odot}$ is the stellar mass, $i$ is the disk inclination angle, and FWHM is with respect to the model component (SC, NC or BC) (see, e.g., \citealt{Salyk_2011, France_2012}).  Kinematic broadening is expected to dominate fluorescent H\textsubscript{2} emission because substantial thermal broadening would require temperatures higher than the dissociation temperature of H\textsubscript{2} (\citealt{Lepp_1983}).  The assumption of kinematic broadening has also been applied to analyses of [O I] 6300 {\AA} (e.g., \citealt{Simon_2016, Fang_2018, Gangi_2020, Campbell-White_2023a}).

\section{Results} \label{sec:4}

We summarize the detection statistics of our fitting routine in Table \ref{table:3}.  Of the 36 CTTS disks in our sample, we detect the [1,4] progression of H\textsubscript{2} above the S/N threshold in all 36 disks (as guaranteed by the use of the (1–7) R(3) emission line of the [1,4] progression to make the cut of the original list of 91 CTTSs).  This proportion falls to 23 of 36 disks in [0,2] emission, and 15 of 36 in [3,16].  This is expected as we use just a single emission line to fit [0,2], and the integrated flux from the major lines of [3,16] is less than the flux from the other progressions in a majority of T Tauri stars (see, e.g., \citealt{France_2012}).  Approximately 2/3 of the [1,4] and [0,2] line profiles are best fit by two Gaussian components, falling to 1/2 for [1,7].  The majority of [3,16] line profiles are best fit by a single Gaussian component, but we note that the relatively low S/N of the sample may preclude the correct classification of true two-component disks.

\subsection{Overview of Fluorescent H\textsubscript{2} Properties}

\begin{figure}
  \centering
  \includegraphics[width=0.99\linewidth]{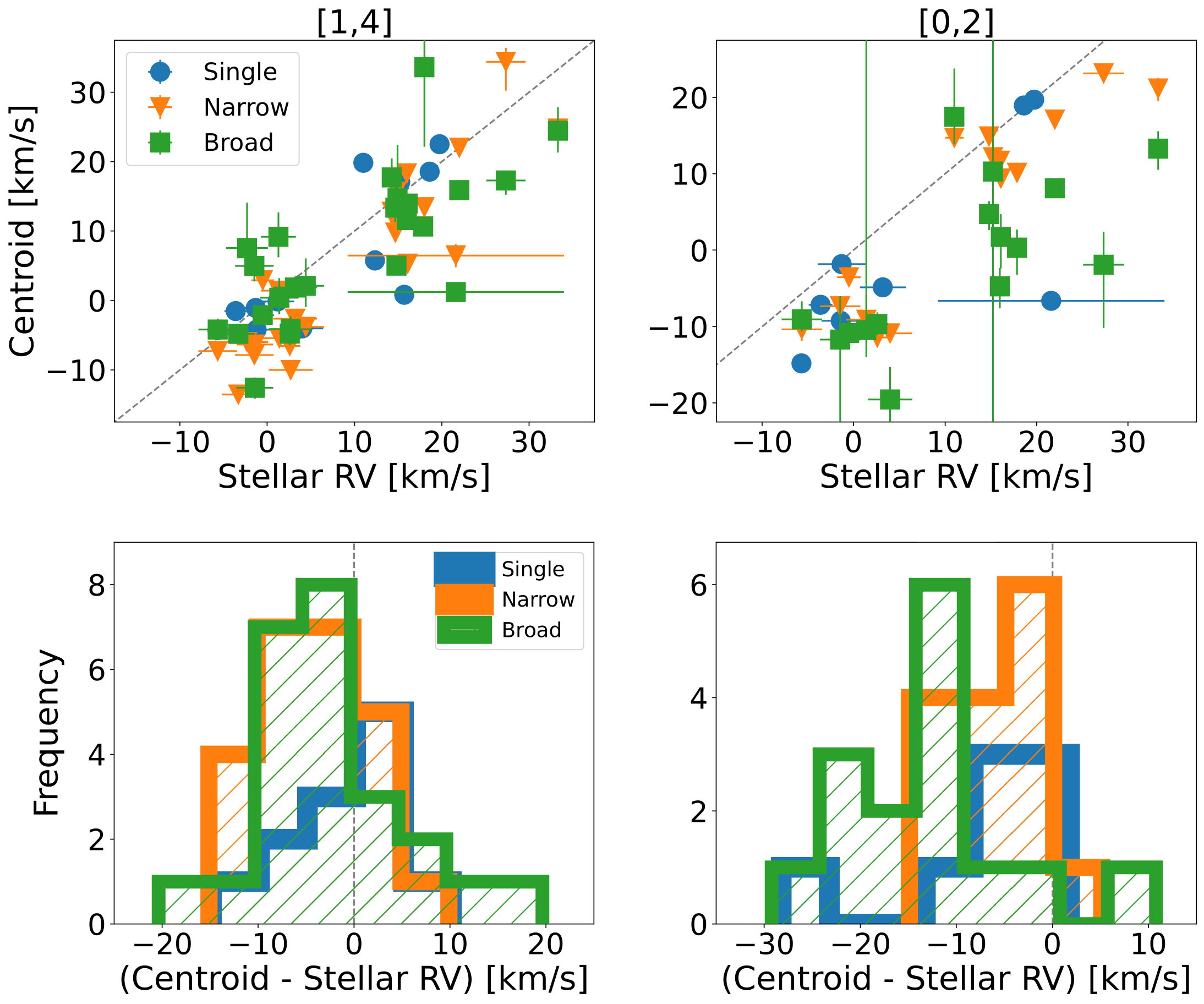}
  \caption{Scatter plots of component centroid velocity and stellar RV (top panels), and histograms of the distribution of RV-corrected component velocities for both fluorescent H\textsubscript{2} progressions (bottom panels).  The markers clustered around zero RV in the top panels consist mostly of disks in the Lupus star-forming region, while markers clustered around 15 km s\textsuperscript{-1} pertain mostly to Taurus-Auriga.}
  \label{fig:4}
\end{figure}

\begin{figure}
  \centering
  \includegraphics[width=0.87\linewidth]{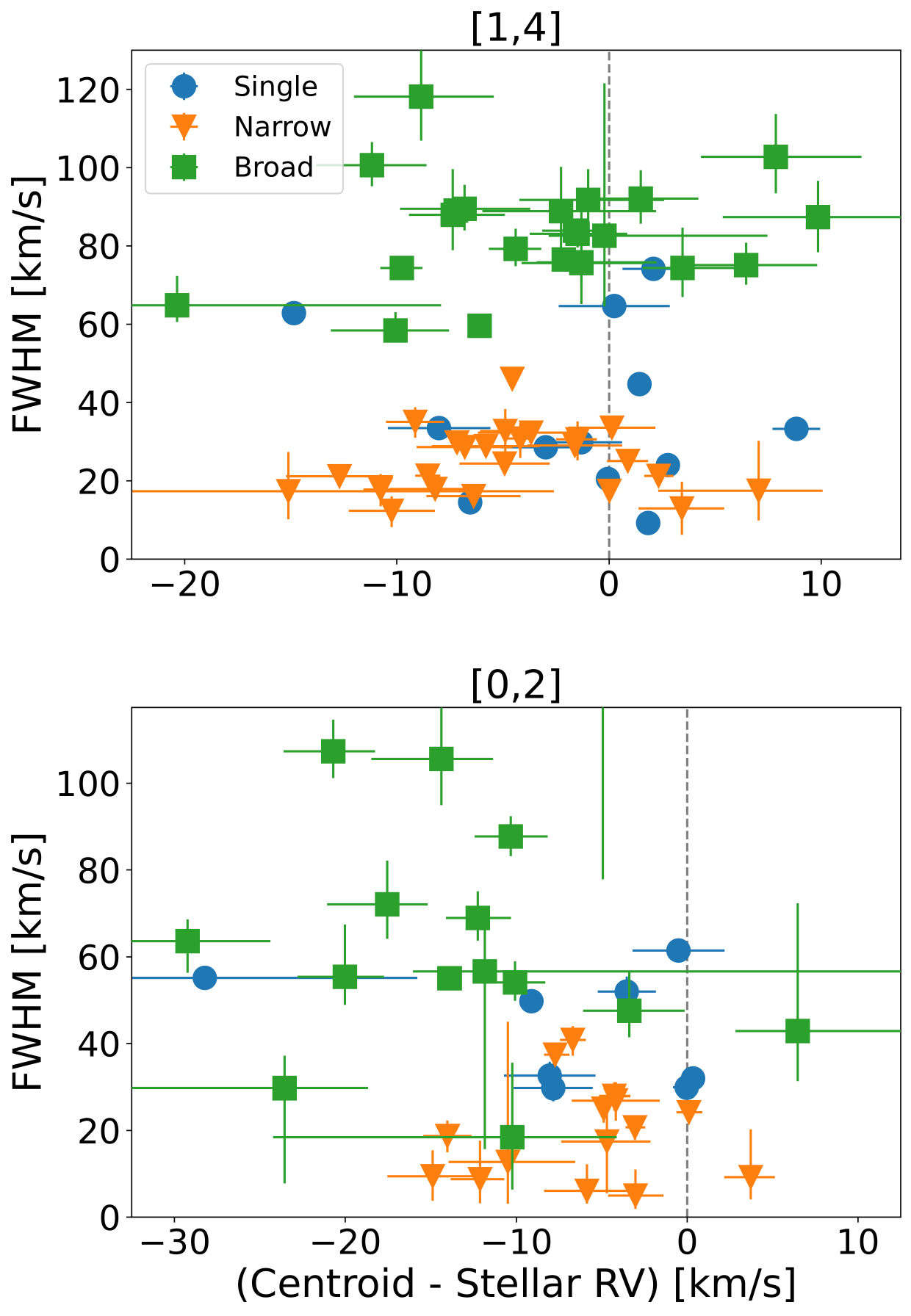}
  \caption{Distribution of FWHM with respect to RV-corrected centroid velocity for both fluorescent progressions.  Component FWHMs in excess of 125 km s\textsuperscript{-1} (BCs of RECX 11 and CS Cha) are excluded for easier visual comparison.}
  \label{fig:5}
\end{figure}

We report the RV-corrected centroid velocity, FWHM and average H\textsubscript{2} emitting radius for each Gaussian component of [1,4] and [0,2] in Table \ref{table:4}.  We visualize the distribution of line kinematics (centroid velocities and FWHMs) in Figures \ref{fig:4} and \ref{fig:5}, respectively.  We exclude properties of the lower-S/N [1,7] emission from further analysis due to similarities with the line kinematics of [1,4] in most sources.  We exclude [3,16] due to its low detection frequency ($<$50\% of disks meet the S/N threshold) and generally messy line profiles (likely due to line overlap with fainter progressions) which complicate the one- or two-component classification of individual disks.  We report the fit parameters of [1,7] and [3,16] in Table \ref{table:7} of Appendix \hyperref[appendix:A]{A}.

\begin{table}
\centering
\textbf{Table 3} \\
\text{Classification of Gaussian Components} \\
\smallskip
 \begin{tabular*}{\linewidth}{c @{\extracolsep{\fill}} ccc}
 \hline
 \hline
 Progression & \# of Disks\textsuperscript{(a)} & SC & NC + BC \\
 \hline
 [1,4] & 36 (100\%) & 12 (33\%) & 24 (67\%) \\
 {[1,7]} & 35 (97\%) & 17 (49\%) & 18 (51\%) \\
 {[0,2]} & 23 (64\%)\textsuperscript{(b)} & 8 (35\%) & 15 (65\%) \\
 {[3,16]} & 15 (42\%) & 11 (73\%) & 4 (27\%) \\
 \hline
 \end{tabular*}
\begin{flushleft}
\textbf{Notes.} (a) Number of sources that meet the S/N requirement for a given progression.  (b) An additional disk, SU Aur, meets the S/N threshold, but is not included due to difficulty in producing clear fits following the MCMC routine.
\end{flushleft}
\rtask{table:3}
\end{table}

\begin{table*}[t!]
\centering
\textbf{Table 4} \\
\text{Properties of the H\textsubscript{2} Emission} \\
\smallskip
 \begin{tabular*}{\textwidth}{c @{\extracolsep{\fill}} cccccccc}
 \hline
 \hline
 & \multicolumn{3}{c}{[1,4]} & & \multicolumn{3}{c}{[0,2]} \\
 \cmidrule(lr){2-4} \cmidrule(lr){6-8}
 Name & Centroid - RV\textsuperscript{(a)} & FWHM\textsuperscript{(b)} & $R$(H\textsubscript{2})\textsuperscript{(c)} & ID\textsuperscript{(d)} & Centroid - RV & FWHM & $R$(H\textsubscript{2}) & ID \\
 & (km s\textsuperscript{-1}) & (km s\textsuperscript{-1}) & (au) & & (km s\textsuperscript{-1}) & (km s\textsuperscript{-1}) & (au) & \\
 \hline
 AA Tau & $-7.17^{+0.15}_{-0.15}$ & $29.80^{+0.72}_{-0.72}$ & $2.35^{+0.11}_{-0.11}$ & NC 
        & $-7.73^{+0.84}_{-0.65}$ & $37.42^{+2.16}_{-2.65}$ & $1.49^{+0.17}_{-0.21}$ & NC \\
        & $-7.22^{+0.52}_{-0.50}$ & $89.04^{+2.27}_{-2.24}$ & $0.26^{+0.01}_{-0.01}$ & BC 
        & $-17.54^{+2.37}_{-3.52}$ & $72.13^{+10.03}_{-7.96}$ & $0.40^{+0.11}_{-0.09}$ & BC \\
 AK Sco & $0.24^{+2.62}_{-2.62}$ & $64.65^{+0.73}_{-0.72}$ & $0.99^{+0.02}_{-0.02}$ & SC 
        & $-0.51^{+2.70}_{-2.69}$ & $61.47^{+1.84}_{-1.78}$ & $1.09^{+0.07}_{-0.06}$ & SC \\
 BP Tau & $-10.76^{+0.87}_{-0.81}$ & $17.78^{+3.95}_{-4.30}$ & $1.35^{+0.60}_{-0.65}$ & NC 
        & $-6.70^{+0.77}_{-0.75}$ & $40.83^{+3.19}_{-3.65}$ & $0.26^{+0.04}_{-0.05}$ & NC \\
        & $-2.14^{+0.37}_{-0.35}$ & $76.53^{+1.16}_{-1.05}$ & $0.07^{+0.00}_{-0.00}$ & BC 
        & $-14.37^{+3.00}_{-4.10}$ & $105.62^{+13.72}_{-10.69}$ & $0.04^{+0.01}_{-0.01}$ & BC \\
 CS Cha & $1.84^{+0.39}_{-0.39}$ & $9.23^{+0.59}_{-0.62}$ & $15.84^{+2.03}_{-2.13}$ & SC 
        & $-3.06^{+0.61}_{-0.55}$ & $20.73^{+1.89}_{-2.89}$ & $3.14^{+0.57}_{-0.88}$ & NC \\
        & $\compactcdots$ & $\compactcdots$ & $\compactcdots$ & $\compactcdots$ 
        & $-4.94^{+45.31}_{-55.66}$ & $761.30^{+2179.76}_{-683.41}$ & ${<}0.01^{+0.01}_{-0.00}$ & BC \\
 DE Tau & $-4.90^{+1.09}_{-1.18}$ & $32.72^{+5.62}_{-7.55}$ & $0.27^{+0.09}_{-0.12}$ & NC 
        & $\compactcdots$ & $\compactcdots$ & $\compactcdots$ & $\compactcdots$ \\
        & $-1.31^{+3.55}_{-2.11}$ & $75.82^{+14.42}_{-10.66}$ & $0.05^{+0.02}_{-0.01}$ & BC 
        & $\compactcdots$ & $\compactcdots$ & $\compactcdots$ & $\compactcdots$ \\
 DF Tau & $0.88^{+0.96}_{-0.98}$ & $25.06^{+2.87}_{-2.78}$ & $2.19^{+0.50}_{-0.49}$ & NC 
        & $0.10^{+0.79}_{-0.73}$ & $24.20^{+1.57}_{-1.90}$ & $2.35^{+0.30}_{-0.37}$ & NC \\
        & $-9.76^{+0.97}_{-1.02}$ & $74.37^{+2.24}_{-2.08}$ & $0.25^{+0.01}_{-0.01}$ & BC 
        & $-10.08^{+1.77}_{-2.18}$ & $54.10^{+4.87}_{-4.24}$ & $0.47^{+0.08}_{-0.07}$ & BC \\
 DM Tau & $-0.06^{+0.21}_{-0.21}$ & $20.47^{+0.72}_{-0.70}$ & $1.39^{+0.10}_{-0.10}$ & SC 
        & $0.34^{+0.52}_{-0.52}$ & $31.94^{+1.45}_{-1.50}$ & $0.57^{+0.05}_{-0.05}$ & SC \\
 ECHA J0843.3 & $0.01^{+0.29}_{-0.28}$ & $17.44^{+1.20}_{-1.24}$ & $1.75^{+0.24}_{-0.25}$ & NC 
              & $-4.89^{+0.96}_{-0.94}$ & $24.95^{+3.15}_{-3.16}$ & $0.86^{+0.22}_{-0.22}$ & NC \\
              & $-6.11^{+0.33}_{-0.34}$ & $59.60^{+1.30}_{-1.21}$ & $0.15^{+0.01}_{-0.01}$ & BC 
              & $-13.90^{+0.72}_{-0.83}$ & $55.06^{+2.01}_{-1.61}$ & $0.18^{+0.01}_{-0.01}$ & BC \\
 GM Aur & $2.33^{+0.70}_{-0.68}$ & $21.26^{+2.39}_{-2.35}$ & $6.27^{+1.41}_{-1.39}$ & NC 
        & $-4.27^{+0.95}_{-0.89}$ & $27.91^{+3.17}_{-3.35}$ & $3.64^{+0.83}_{-0.87}$ & NC \\
        & $-4.41^{+1.21}_{-1.27}$ & $79.26^{+5.14}_{-4.43}$ & $0.45^{+0.06}_{-0.05}$ & BC 
        & $-20.70^{+2.45}_{-2.91}$ & $107.38^{+7.32}_{-6.18}$ & $0.25^{+0.03}_{-0.03}$ & BC \\
 HN Tau & $7.05^{+3.02}_{-4.74}$ & $17.47^{+12.76}_{-7.60}$ & $10.78^{+15.75}_{-9.38}$ & NC 
        & $-4.18^{+2.57}_{-2.58}$ & $26.88^{+4.21}_{-4.61}$ & $4.55^{+1.43}_{-1.56}$ & NC \\
        & $-10.06^{+2.53}_{-3.05}$ & $58.41^{+4.68}_{-2.29}$ & $0.96^{+0.15}_{-0.08}$ & BC 
        & $-29.23^{+4.85}_{-8.60}$ & $63.63^{+5.03}_{-7.30}$ & $0.81^{+0.13}_{-0.19}$ & BC \\
 MY Lup & $-8.19^{+2.15}_{-2.15}$ & $17.92^{+2.10}_{-2.26}$ & $10.59^{+2.48}_{-2.67}$ & NC 
        & $\compactcdots$ & $\compactcdots$ & $\compactcdots$ & $\compactcdots$ \\
        & $-2.27^{+4.48}_{-3.71}$ & $88.88^{+11.31}_{-10.21}$ & $0.43^{+0.11}_{-0.10}$ & BC 
        & $\compactcdots$ & $\compactcdots$ & $\compactcdots$ & $\compactcdots$ \\
 RECX 11 & $-4.57^{+0.35}_{-0.34}$ & $46.04^{+1.46}_{-1.56}$ & $1.23^{+0.08}_{-0.08}$ & NC 
         & $\compactcdots$ & $\compactcdots$ & $\compactcdots$ & $\compactcdots$ \\
         & $15.61^{+25.82}_{-11.47}$ & $204.33^{+75.10}_{-37.20}$ & $0.06^{+0.05}_{-0.02}$ & BC 
         & $\compactcdots$ & $\compactcdots$ & $\compactcdots$ & $\compactcdots$ \\
 RX J1556.1 & $-9.13^{+1.36}_{-1.38}$ & $35.06^{+3.74}_{-4.05}$ & $0.87^{+0.18}_{-0.20}$ & NC 
            & $-14.02^{+1.39}_{-1.41}$ & $18.73^{+3.59}_{-3.77}$ & $3.03^{+1.16}_{-1.22}$ & NC \\
            & $-7.36^{+2.44}_{-2.06}$ & $87.95^{+11.68}_{-9.03}$ & $0.14^{+0.04}_{-0.03}$ & BC 
            & $-12.26^{+1.94}_{-1.84}$ & $68.99^{+6.15}_{-5.25}$ & $0.22^{+0.04}_{-0.03}$ & BC \\
 RX J1842.9 & $-1.62^{+2.23}_{-2.23}$ & $29.03^{+1.84}_{-1.87}$ & $1.35^{+0.17}_{-0.17}$ & NC 
            & $-4.70^{+2.55}_{-2.68}$ & $17.45^{+10.48}_{-11.95}$ & $3.73^{+4.48}_{-3.73}$ & NC \\
            & $1.49^{+2.72}_{-2.56}$ & $92.09^{+7.25}_{-6.42}$ & $0.13^{+0.02}_{-0.02}$ & BC 
            & $-3.39^{+3.24}_{-2.71}$ & $47.57^{+9.26}_{-6.15}$ & $0.50^{+0.20}_{-0.13}$ & BC \\
 RX J1852.3 & $-6.39^{+2.21}_{-2.21}$ & $16.09^{+0.62}_{-0.63}$ & $3.60^{+0.28}_{-0.28}$ & NC 
            & $-5.87^{+2.48}_{-2.51}$ & $6.07^{+6.14}_{-2.95}$ & $25.28^{+51.15}_{-24.57}$ & NC \\
            & $6.46^{+3.35}_{-3.15}$ & $75.11^{+5.72}_{-5.01}$ & $0.17^{+0.03}_{-0.02}$ & BC 
            & $-10.24^{+6.11}_{-13.99}$ & $18.45^{+17.12}_{-12.09}$ & $2.74^{+5.08}_{-2.74}$ & BC \\
 RY Lup & $0.13^{+2.05}_{-2.05}$ & $33.57^{+1.71}_{-1.84}$ & $4.14^{+0.42}_{-0.45}$ & NC 
        & $\compactcdots$ & $\compactcdots$ & $\compactcdots$ & $\compactcdots$ \\
        & $7.86^{+4.04}_{-3.53}$ & $102.76^{+10.96}_{-9.33}$ & $0.44^{+0.09}_{-0.08}$ & BC 
        & $\compactcdots$ & $\compactcdots$ & $\compactcdots$ & $\compactcdots$ \\
 SSTc2d J160830.7 & $-1.32^{+1.93}_{-1.93}$ & $29.84^{+0.92}_{-0.91}$ & $5.52^{+0.34}_{-0.34}$ & SC 
                  & $\compactcdots$ & $\compactcdots$ & $\compactcdots$ & $\compactcdots$ \\
 SSTc2d J161243.8 & $-3.68^{+2.51}_{-2.51}$ & $32.31^{+3.02}_{-2.68}$ & $0.73^{+0.14}_{-0.12}$ & NC 
                  & $\compactcdots$ & $\compactcdots$ & $\compactcdots$ & $\compactcdots$ \\
                  & $9.85^{+6.96}_{-4.49}$ & $87.36^{+9.28}_{-8.97}$ & $0.10^{+0.02}_{-0.02}$ & BC 
                  & $\compactcdots$ & $\compactcdots$ & $\compactcdots$ & $\compactcdots$ \\
 SU Aur & $-1.49^{+0.91}_{-1.01}$ & $30.60^{+4.59}_{-5.37}$ & $6.56^{+1.97}_{-2.30}$ & NC 
        & $\compactcdots$ & $\compactcdots$ & $\compactcdots$ & $\compactcdots$ \\
        & $3.46^{+2.75}_{-1.88}$ & $74.38^{+10.31}_{-7.44}$ & $1.11^{+0.31}_{-0.22}$ & BC 
        & $\compactcdots$ & $\compactcdots$ & $\compactcdots$ & $\compactcdots$ \\
 SY Cha & $-14.85^{+0.44}_{-0.43}$ & $62.84^{+0.86}_{-0.86}$ & $0.42^{+0.01}_{-0.01}$ & SC 
        & $\compactcdots$ & $\compactcdots$ & $\compactcdots$ & $\compactcdots$ \\
 Sz 45 & $-4.19^{+0.54}_{-0.65}$ & $30.80^{+3.94}_{-4.97}$ & $0.97^{+0.25}_{-0.31}$ & NC 
       & $\compactcdots$ & $\compactcdots$ & $\compactcdots$ & $\compactcdots$ \\
       & $-0.22^{+7.69}_{-2.63}$ & $82.61^{+38.94}_{-18.02}$ & $0.14^{+0.13}_{-0.06}$ & BC 
       & $\compactcdots$ & $\compactcdots$ & $\compactcdots$ & $\compactcdots$ \\
 Sz 71 & $-10.24^{+2.04}_{-2.02}$ & $12.35^{+3.69}_{-4.16}$ & $4.04^{+2.41}_{-2.72}$ & NC 
       & $\compactcdots$ & $\compactcdots$ & $\compactcdots$ & $\compactcdots$ \\
       & $-1.49^{+2.34}_{-2.26}$ & $83.10^{+3.93}_{-3.51}$ & $0.09^{+0.01}_{-0.01}$ & BC 
       & $\compactcdots$ & $\compactcdots$ & $\compactcdots$ & $\compactcdots$ \\
 Sz 75 & $2.09^{+1.45}_{-1.46}$ & $74.14^{+1.83}_{-1.79}$ & $0.43^{+0.02}_{-0.02}$ & SC 
       & $-3.54^{+1.71}_{-1.70}$ & $51.99^{+3.48}_{-3.43}$ & $0.88^{+0.12}_{-0.12}$ & SC \\
 Sz 82 & $3.43^{+1.99}_{-2.05}$ & $12.95^{+6.83}_{-6.74}$ & $13.00^{+13.71}_{-13.00}$ & NC 
       & $-3.03^{+1.65}_{-1.60}$ & $4.98^{+6.01}_{-3.10}$ & $87.89^{+212.13}_{-87.89}$ & NC \\
       & $-1.61^{+1.55}_{-1.55}$ & $83.90^{+2.59}_{-2.33}$ & $0.31^{+0.02}_{-0.02}$ & BC 
       & $-10.33^{+2.16}_{-2.12}$ & $87.72^{+4.69}_{-4.55}$ & $0.28^{+0.03}_{-0.03}$ & BC \\
 Sz 98 & $-4.93^{+2.13}_{-2.13}$ & $24.41^{+1.62}_{-1.72}$ & $2.28^{+0.30}_{-0.32}$ & NC 
       & $-7.84^{+2.31}_{-2.32}$ & $29.74^{+3.19}_{-3.02}$ & $1.45^{+0.31}_{-0.30}$ & SC \\
       & $-11.17^{+2.57}_{-2.63}$ & $100.63^{+5.90}_{-5.38}$ & $0.13^{+0.02}_{-0.01}$ & BC 
       & $\compactcdots$ & $\compactcdots$ & $\compactcdots$ & $\compactcdots$ \\
 Sz 100 & $-12.70^{+2.52}_{-2.52}$ & $21.16^{+1.59}_{-1.58}$ & $0.59^{+0.09}_{-0.09}$ & NC 
        & $\compactcdots$ & $\compactcdots$ & $\compactcdots$ & $\compactcdots$ \\
        & $-6.82^{+3.10}_{-3.02}$ & $89.49^{+6.14}_{-5.53}$ & $0.03^{+0.00}_{-0.00}$ & BC 
        & $\compactcdots$ & $\compactcdots$ & $\compactcdots$ & $\compactcdots$ \\
 \end{tabular*}
\rtask{table:4}
\end{table*}

\begin{table*}
\centering
\textbf{Table 4} \\
\text{(Continued)} \\
\smallskip
 \begin{tabular*}{\textwidth}{c @{\extracolsep{\fill}} cccccccc}
 \hline
 \hline
 & \multicolumn{3}{c}{[1,4]} & & \multicolumn{3}{c}{[0,2]} \\
 \cmidrule(lr){2-4} \cmidrule(lr){6-8}
 Name & Centroid - RV & FWHM & $R$(H\textsubscript{2}) & ID & Centroid - RV & FWHM & $R$(H\textsubscript{2}) & ID \\
 & (km s\textsuperscript{-1}) & (km s\textsuperscript{-1}) & (au) & & (km s\textsuperscript{-1}) & (km s\textsuperscript{-1}) & (au) & \\
 \hline
 Sz 102 & $-15.11^{+12.50}_{-12.51}$ & $17.34^{+10.03}_{-7.16}$ & $1.81^{+2.09}_{-1.49}$ & NC 
        & $-28.22^{+12.44}_{-12.44}$ & $55.15^{+2.28}_{-2.22}$ & $0.18^{+0.01}_{-0.01}$ & SC \\
        & $-20.36^{+12.45}_{-12.47}$ & $64.84^{+7.50}_{-4.26}$ & $0.13^{+0.03}_{-0.02}$ & BC 
        & $\compactcdots$ & $\compactcdots$ & $\compactcdots$ & $\compactcdots$ \\
 Sz 103 & $-6.81^{+2.25}_{-2.25}$ & $28.62^{+1.81}_{-1.96}$ & $0.61^{+0.08}_{-0.08}$ & NC 
        & $-10.50^{+3.94}_{-3.45}$ & $12.74^{+32.28}_{-9.65}$ & $3.08^{+15.60}_{-3.08}$ & NC \\
        & $-0.99^{+3.58}_{-3.25}$ & $91.76^{+7.85}_{-7.35}$ & $0.06^{+0.01}_{-0.01}$ & BC 
        & $-11.85^{+63.61}_{-4.18}$ & $56.64^{+15.09}_{-40.93}$ & $0.16^{+0.08}_{-0.23}$ & BC \\
 Sz 111 & $-2.99^{+2.12}_{-2.12}$ & $28.57^{+0.84}_{-0.82}$ & $1.28^{+0.08}_{-0.07}$ & SC 
        & $\compactcdots$ & $\compactcdots$ & $\compactcdots$ & $\compactcdots$ \\
 Sz 114 & $-8.01^{+2.41}_{-2.41}$ & $33.48^{+0.73}_{-0.73}$ & $0.01^{+0.00}_{-0.00}$ & SC 
        & $-14.88^{+2.70}_{-2.65}$ & $9.43^{+6.01}_{-5.68}$ & $0.10^{+0.12}_{-0.12}$ & NC \\
        & $\compactcdots$ & $\compactcdots$ & $\compactcdots$ & $\compactcdots$ 
        & $-23.55^{+4.89}_{-15.88}$ & $29.74^{+7.44}_{-21.96}$ & $0.01^{+0.00}_{-0.01}$ & BC \\
 Sz 129 & $-5.81^{+2.54}_{-2.55}$ & $28.85^{+2.38}_{-2.44}$ & $0.93^{+0.15}_{-0.16}$ & NC 
        & $-8.05^{+2.68}_{-2.68}$ & $32.59^{+3.20}_{-2.97}$ & $0.73^{+0.14}_{-0.13}$ & SC \\
        & $-1.31^{+2.96}_{-2.82}$ & $75.63^{+5.61}_{-5.03}$ & $0.13^{+0.02}_{-0.02}$ & BC 
        & $\compactcdots$ & $\compactcdots$ & $\compactcdots$ & $\compactcdots$ \\
 TW Hya & $-6.56^{+0.03}_{-0.03}$ & $14.50^{+0.14}_{-0.14}$ & $0.05^{+0.00}_{-0.00}$ & SC 
        & $\compactcdots$ & $\compactcdots$ & $\compactcdots$ & $\compactcdots$ \\
 UX Tau A & $2.77^{+0.39}_{-0.39}$ & $24.01^{+0.57}_{-0.58}$ & $2.75^{+0.13}_{-0.13}$ & SC 
          & $-0.03^{+0.80}_{-0.81}$ & $29.88^{+2.36}_{-2.31}$ & $1.78^{+0.28}_{-0.27}$ & SC \\
 V4046 Sgr & $1.43^{+0.23}_{-0.23}$ & $44.71^{+0.22}_{-0.22}$ & $0.53^{+0.01}_{-0.01}$ & SC 
           & $-9.12^{+0.30}_{-0.30}$ & $49.77^{+0.48}_{-0.47}$ & $0.43^{+0.01}_{-0.01}$ & SC \\
 V510 Ori & $-8.54^{+0.59}_{-0.59}$ & $21.31^{+2.44}_{-2.57}$ & $\compactcdots$ & NC 
          & $-12.13^{+1.42}_{-1.70}$ & $8.76^{+8.90}_{-5.56}$ & $\compactcdots$ & NC \\
          & $-8.85^{+3.41}_{-3.17}$ & $118.15^{+12.97}_{-11.26}$ & $\compactcdots$ & BC 
          & $-20.02^{+2.29}_{-2.78}$ & $55.40^{+12.08}_{-6.48}$ & $\compactcdots$ & BC \\
 VZ Cha & $8.83^{+1.12}_{-1.13}$ & $33.27^{+1.75}_{-1.74}$ & $0.46^{+0.05}_{-0.05}$ & SC 
        & $3.72^{+1.41}_{-1.55}$ & $9.23^{+11.03}_{-5.14}$ & $5.95^{+14.22}_{-5.95}$ & NC \\
        & $\compactcdots$ & $\compactcdots$ & $\compactcdots$ & $\compactcdots$ 
        & $6.47^{+6.41}_{-3.65}$ & $42.88^{+29.50}_{-11.58}$ & $0.28^{+0.38}_{-0.15}$ & BC \\
 \hline
 \end{tabular*}
\begin{flushleft}
\textbf{Notes.} (a) Observed centroid velocities corrected for literature values of stellar radial velocity (RV).  Errors from the fitting routine and RVs propagated through.  (b) Per the fitting routine, the FWHM is corrected for instrumental broadening introduced by the COS LSF.  (c) The emitting radius of the H\textsubscript{2}, accounting only for errors on the FWHM values.  (d) Gaussian component type: single (SC), narrow (NC), or broad (BC).
\end{flushleft}
\end{table*}

In terms of RV-corrected centroid velocities, five of the six component distributions peak in the blue, suggestive of a disk-wind origin for fluorescent H\textsubscript{2}.  The median velocities of the SC centroids of [0,2], and the NC centroids of both progressions, peak near -5 km s\textsuperscript{-1}, larger than the wavelength uncertainty of \textit{HST}-COS.  The BC distributions are also blueshifted, substantially so for [0,2] (median value of -12 km s\textsuperscript{-1}).  The SC distribution of [1,4] is consistent with the distribution of stellar radial velocities.

On the level of individual disks, we note a maximum blueshifted velocity of -20 km s\textsuperscript{-1} for the BC of Sz 102 as traced by the [1,4] progression, increasing to -28 km s\textsuperscript{-1} for a SC fit traced by [0,2] in the same disk.  Redshifted peaks exist for at least one component in 17/36 disks for [1,4] and 3/23 disks for [0,2], but it is mostly within the measurement uncertainties, or the wavelength uncertainty of \textit{HST}-COS.  Interestingly, component centroid velocities of the [0,2] progression are consistently blueshifted a few km s\textsuperscript{-1} with respect to velocities of [1,4].  We discuss two possibilities for this difference in Section \ref{sec:6}.

In terms of FWHM, there is clear separation between the BC (median of 84 km s\textsuperscript{-1}) and NC (median of 25 km s\textsuperscript{-1}) distributions of [1,4], somewhat expected given the fitting procedure.  The SC distribution mostly follows the NCs, with a few disks extending into the size range of the BCs.  The distributions are more muddled in [0,2], with narrower median widths than [1,4] for both the broad and narrow components.  Large uncertainties exist on low-amplitude BCs (see, e.g., CS Cha) that may partially fit extraneous continuum features, but the effect of these uncertainties is reduced in the aggregate by comparing sample medians rather than means.  There is no obvious correlation between FWHM and centroid velocity in the components of either distribution.

With respect to inferred H\textsubscript{2} emitting radii, the SC emission of both progressions appears to originate mostly between $\sim$0.5 and 3 au of the central star.  Most BC emission originates interior to 0.5 au, with only a single component inferring emission from beyond 1 au in each progression.  Median emitting radii are approximately 2 and 3 au for the NCs of [1,4] and [0,2], respectively.

\subsection{Correlations of NC and BC Kinematics}

\begin{figure}
  \centering
  \includegraphics[width=\linewidth]{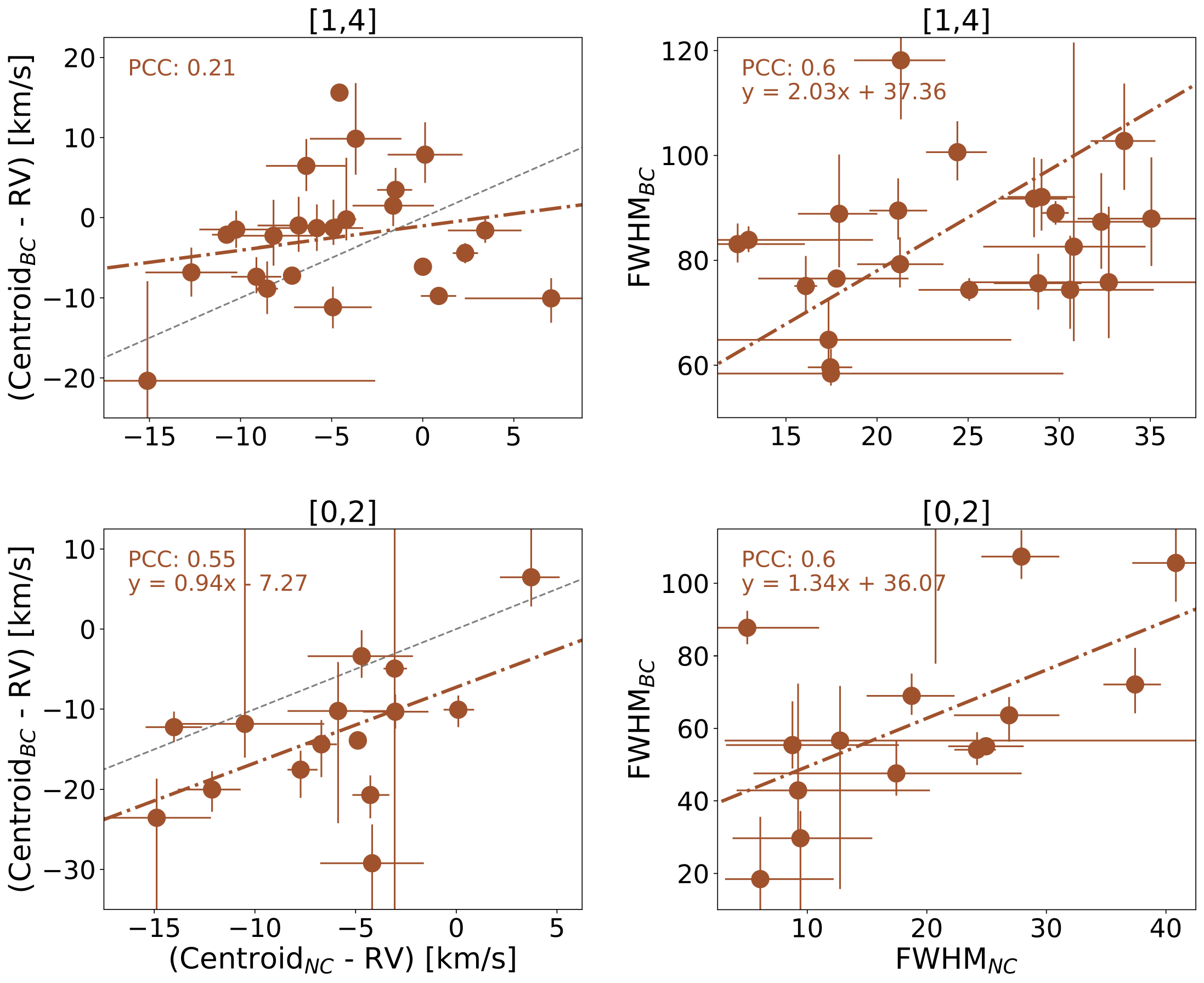}
  \caption{Correlations between the NC and BC kinematic properties for both progressions of fluorescent H\textsubscript{2}.  We report the equation of the line (brown dashes and dots) which best fits the data for statistically-significant correlation pairs.  The light gray dotted line in the velocity plots (left panels) represents a hypothetical one-to-one correlation.}
  \label{fig:6}
\end{figure}

We show correlations between the NC and BC kinematics for two-component disks in Figure \ref{fig:6}.  We use the Pearson correlation coefficient (PCC) to evaluate the statistical significance of a given correlation.  At the 2$\sigma$ confidence level for a two-tailed test of 24 pairs, the critical value of the PCC is 0.40.  For a two-tailed test of 15 pairs, it is 0.51.  For the [1,4] progression, the component FWHMs are correlated with each other above the relevant critical value.  That is, the correlation is statistically significant.  For the [0,2] progression, both the RV-corrected component centroid velocities and component FWHMs are correlated to a statistically-significant degree.  We briefly discuss the potential implication of these correlations in Section \ref{sec:6}, noting here that testing multiple correlations simultaneously may reduce the confidence of any one correlation proportionate to the actual number of correlations tested (\citealt{Miller_1981}).

\subsection{Correlations of Component Properties with Disk Inclination and the Infrared Index}

\begin{figure}
  \centering
  \includegraphics[width=\linewidth]{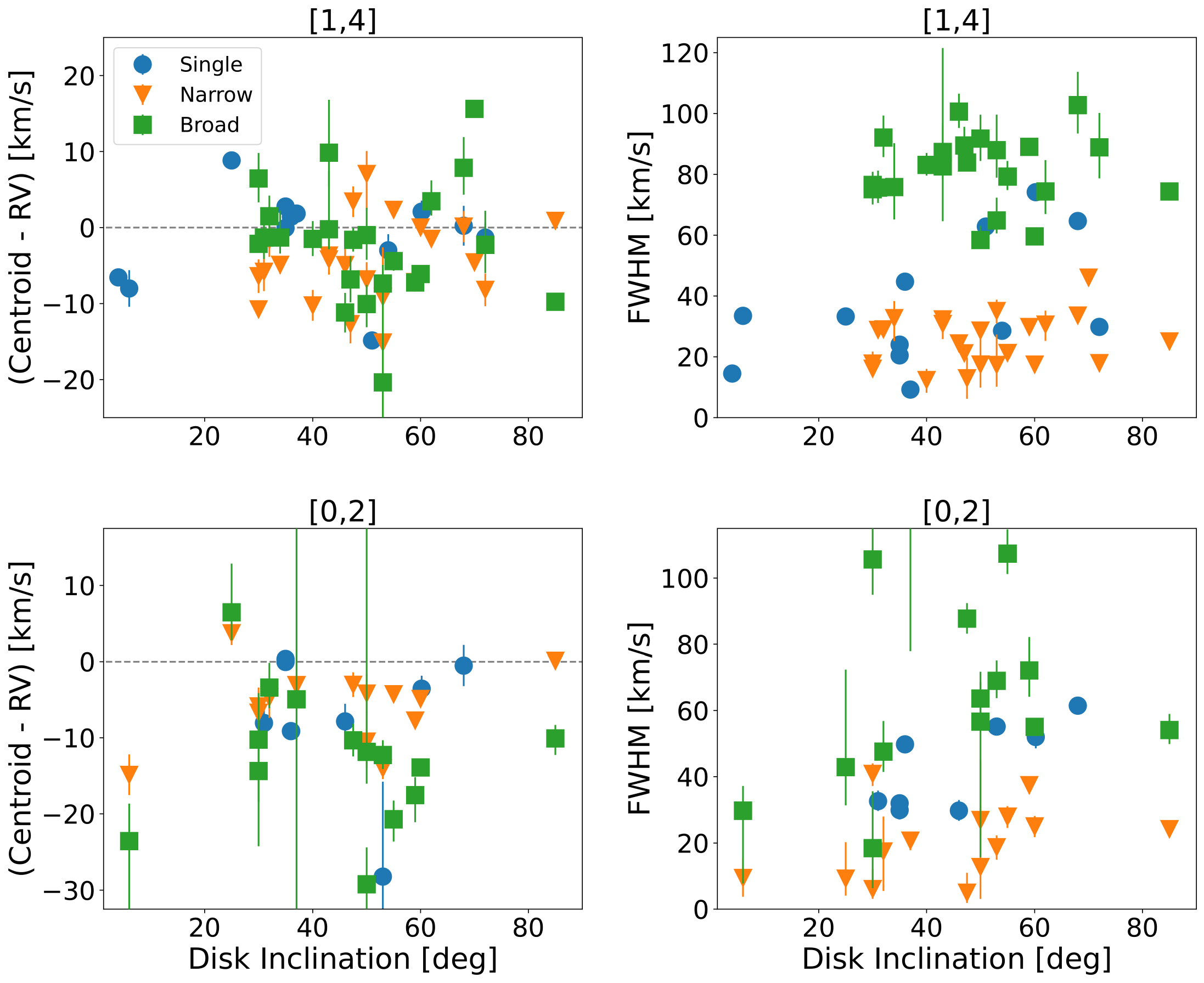}
  \caption{Correlations between H\textsubscript{2} line kinematics and disk inclination for [1,4] and [0,2].  We mark the RV-corrected component velocity zero point with a light gray dotted line (left panels).}
  \label{fig:7}
\end{figure}

In Figure \ref{fig:7}, we compare centroid velocities and FWHMs with disk inclination.  We do not detect linear correlations above the PCC critical values for either the NCs or BCs of [1,4] and [0,2].  We do, however, note that there may be a slight preference for higher blueshifts at disk inclination angles near 50 degrees, particularly for the BCs.  We discuss this finding in Section \ref{sec:6}.

The FWHM SCs show a trend of positive correlation with disk inclination, but this correlation only reaches 2$\sigma$ significance in the eight-disk sample of [0,2] (PCC = 0.78).  This finding is also reported in [O I] 6300 {\AA} emission for SC disks of \citet{Banzatti_2019}.

\begin{figure}
  \centering
  \includegraphics[width=\linewidth]{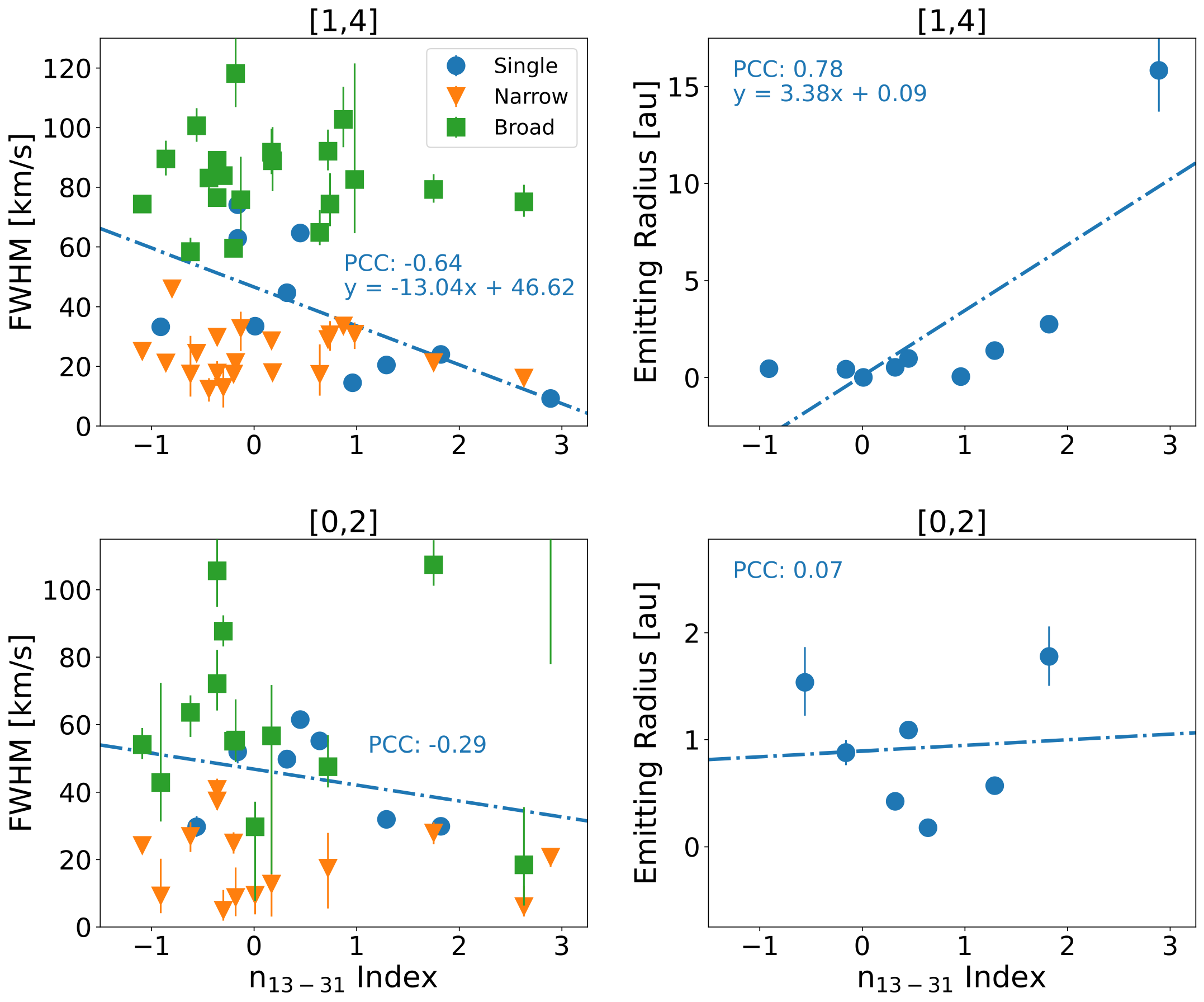}
  \caption{Correlations between component line broadening and the infrared index.  We calculate the Pearson coefficient for the SC distribution in each plot.  We report the equation of the best-fit line (blue dashes and dots) for statistically-significant SC correlations.  We do not plot the NC and BC distributions in the right panels because their emitting radii do not correlate strongly with the infrared index, and the scale of the y-axis is not favorable to visual inspection of small BC emitting radii.}
  \label{fig:8}
\end{figure}

In Figure \ref{fig:8}, we compare component FWHMs and emitting radii with the $n$\textsubscript{13-31} infrared index.  We do not detect significant correlations for the NC or BC distributions in either progression, but see a trend with the SCs of [1,4].  As the $n$\textsubscript{13-31} index increases (i.e., as the inner disk becomes optically thin), SC emitting radii increase.  We do not see a similar trend in [0,2], although the sample size is limited and does not contain disks at either $n$\textsubscript{13-31} extreme (optically thick or optically thin).  While the [1,4] result is supportive of a scenario where disk winds co-evolve with the dusty disk, we do not discuss it further due to the null result in [0,2] (see \citealt{Banzatti_2019} for a discussion pertaining to this correlation observed in [O I] emission).  We note that while removal of the data point near 15 au in the upper-right panel significantly decreases the slope of the best-fit line, it does not substantially change the correlation (PCC = 0.71).

\section{Comparison of H\textsubscript{2} and [O I] 6300 {\AA} Line Properties} \label{sec:5}

\begin{figure}
  \centering
  \includegraphics[width=\linewidth]{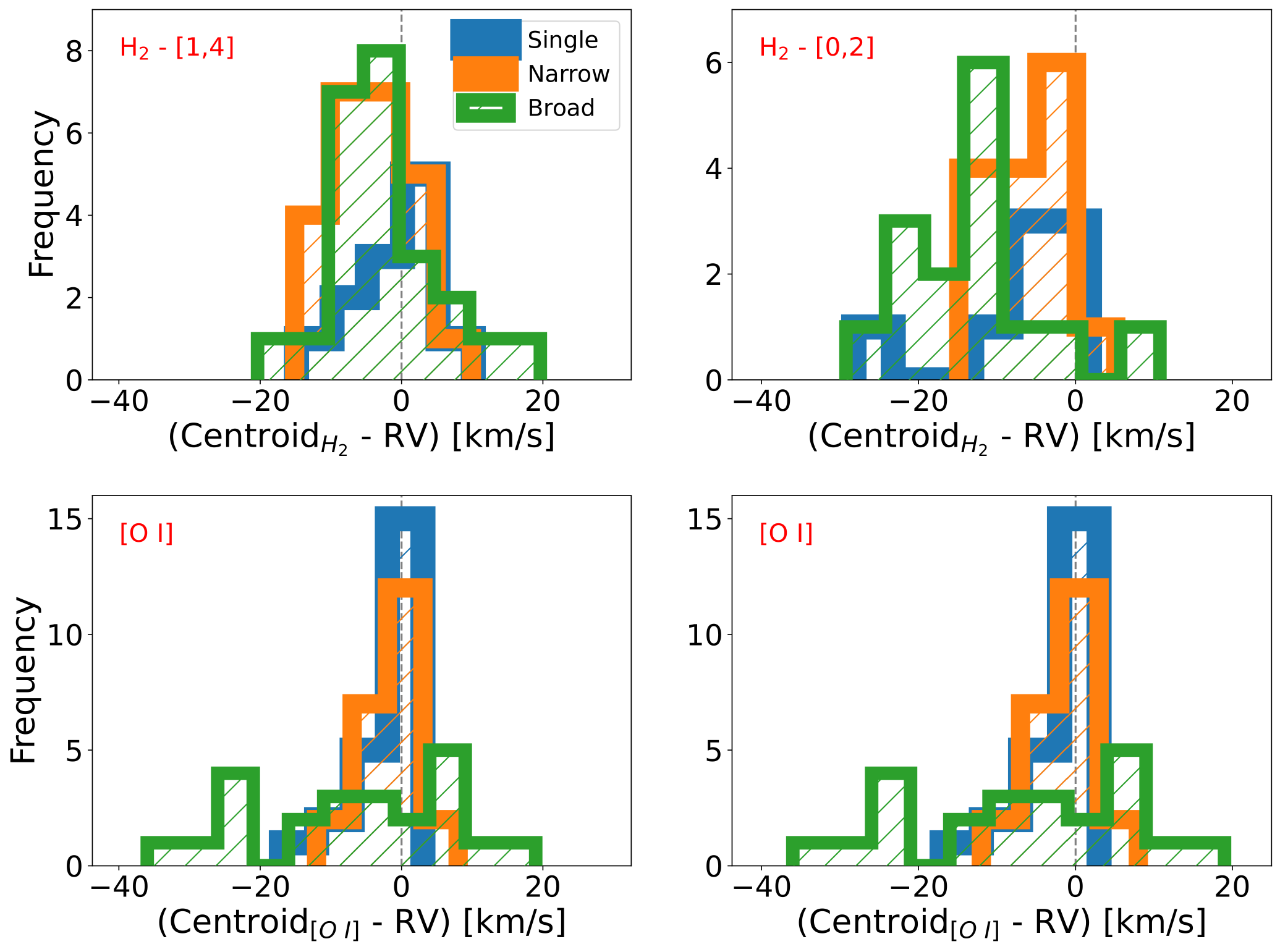}
  \caption{Component velocity comparison of H\textsubscript{2} - [1,4] (left panels) and H\textsubscript{2} - [0,2] (right panels) with [O I] 6300 {\AA}.  SCs are in blue, NCs in orange, and BCs in green.  Zero velocity is marked by a vertical dashed line.  The upper panels are copied from Figure \ref{fig:4} and the lower panels are identical to facilitate comparison.}
  \label{fig:9}
\end{figure}

\begin{figure}
  \centering
  \includegraphics[width=0.87\linewidth]{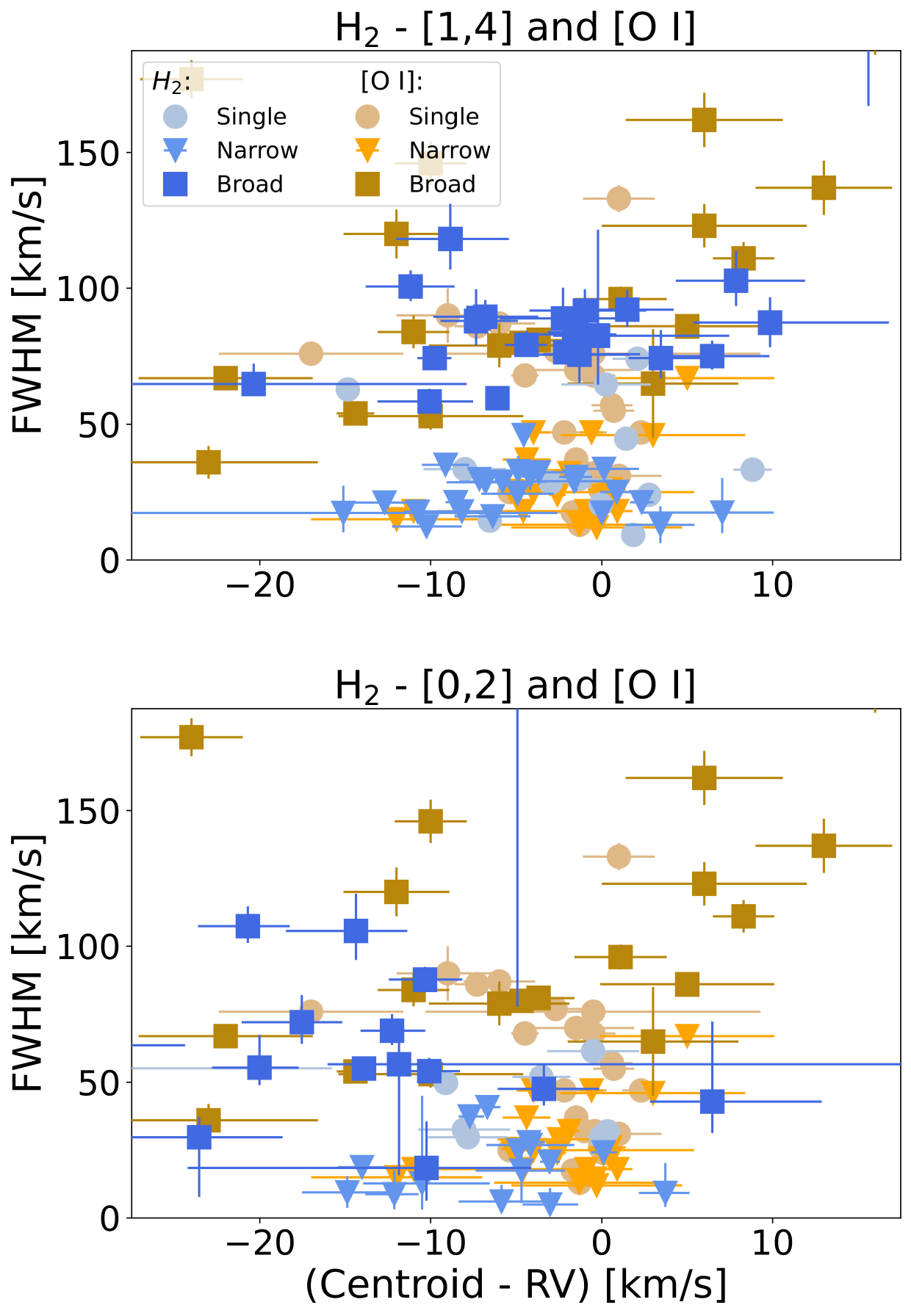}
  \caption{Distribution of component FWHMs with respect to RV-corrected component velocities.  The [1,4] progression of fluorescent H\textsubscript{2} is compared to [O I] 6300 {\AA} in the top panel, and the [0,2] progression is compared to [O I] in the bottom panel.  Component velocities less than -25 km s\textsuperscript{-1}, and FWHMs in excess of 180 km s\textsuperscript{-1}, are cut off from the plot for easier visual comparison.}
  \label{fig:10}
\end{figure}

To better place our kinematic analysis of fluorescent H\textsubscript{2} in the context of existing tracers of low-velocity disk winds, we compare our results to those from the widely-studied [O I] 6300 {\AA} emission line.  In the aggregate, we compare our sample to the sample of 65 T Tauri stars analyzed at a spectral resolution of $\sim$7 km s\textsuperscript{-1} by \citet{Banzatti_2019}.  They self-consistently correct centroid velocities for stellar RV through cross-correlation of the 6439.07 {\AA} photospheric line of calcium with stellar models.  They fit Gaussian components to the [O I] line profiles via $\chi^{2}$ minimization with the IDL routine \textit{MPFIT} (\citealt{Markwardt_2009}), and adopt the same classification scheme that we use here to identify single, narrow and broad components.  Of their 58 disks with identified low-velocity ($<$30 km s\textsuperscript{-1}) [O I] emission, they classify 23 as two-component (NC and BC) disks, and 23 as SC disks without evidence of a high-velocity jet.  An additional 12 SC disks show evidence of a jet (denoted SCJ), which is not seen in fluorescent H\textsubscript{2} emission and which we exclude from our comparison.  We report the median RV-corrected component velocities, FWHMs and emitting radii for the H\textsubscript{2} - [1,4], H\textsubscript{2} - [0,2], and [O I] 6300 {\AA} ensembles in Table \ref{table:5}.  We visualize the comparison of line kinematics in Figures \ref{fig:9} and \ref{fig:10}.

\begin{table*}
\centering
\textbf{Table 5} \\
\text{Comparison of H\textsubscript{2} and [O I] Sample Medians} \\
\smallskip
 \begin{tabular*}{\textwidth}{c @{\extracolsep{\fill}} cccccc}
 \hline
 \hline
 & H\textsubscript{2} - [1,4] & H\textsubscript{2} - [0,2] & [O I] 6300 {\AA}\textsuperscript{(a)} \\
 & (km s\textsuperscript{-1}) & (km s\textsuperscript{-1}) & (km s\textsuperscript{-1}) \\
 \hline
 Centroid\textsubscript{SC} - Stellar RV & $0.09^{+1.34}_{-1.47}$ & $-5.69^{+3.66}_{-2.36}$ & $-1.3^{+0.8}_{-0.2}$ \\
 Centroid\textsubscript{NC} - Stellar RV & $-4.92^{+0.72}_{-1.18}$ & $-4.89^{+0.62}_{-1.81}$ & $-0.9^{+0.5}_{-1.0}$ \\
 Centroid\textsubscript{BC} - Stellar RV & $-1.88^{+0.56}_{-2.31}$ & $-12.26^{+1.93}_{-2.11}$ & $-6.0^{+2.3}_{-4.0}$ \\
 \hline
 FWHM\textsubscript{SC} & $31.56^{+7.44}_{-2.98}$ & $41.18^{+9.70}_{-9.24}$ & $55.0^{+13.0}_{-8.0}$ \\
 FWHM\textsubscript{NC} & $24.74^{+4.00}_{-3.47}$ & $18.73^{+5.47}_{-5.99}$ & $25.0^{+1.0}_{-1.0}$ \\
 FWHM\textsubscript{BC} & $83.50^{+4.45}_{-4.24}$ & $56.64^{+12.35}_{-1.58}$ & $86.0^{+25.0}_{-5.0}$ \\
 \hline
 & (au) & (au) & (au) \\
 Radius\textsubscript{SC} & $0.76^{+0.52}_{-0.30}$ & $0.80^{+0.29}_{-0.23}$ & $0.5^{+0.1}_{-0.2}$ \\
 Radius\textsubscript{NC} & $1.81^{+0.47}_{-0.46}$ & $3.11^{+0.53}_{-0.42}$ & $2.5^{+0.4}_{-1.5}$ \\
 Radius\textsubscript{BC} & $0.14^{+0.01}_{-0.00}$ & $0.26^{+0.08}_{-0.05}$ & $0.1^{+0.1}_{-0.1}$ \\
 \hline
 & (p-value) & (p-value) & (p-value) \\
 K-S Test 1 (RVs x Centroids) & 0.17 & 0.02 & 0.13 \\
 K-S Test 2 (H\textsubscript{2} x [O I]) & 0.05 & 0.01 & $\compactcdots$ \\
 \hline
 \end{tabular*}
\begin{flushleft}
\textbf{Notes:} The reported errors represent the one-sigma confidence interval of the sample medians.  They are calculated by bootstrapping with scipy.stats.bootstrap. The first Kolmogorov–Smirnov (K-S) test asks if the stellar RVs (one- and two-component disks) and component centroids (single, narrow and broad) are sampled from the same distribution. The second K-S test asks if the H\textsubscript{2} centroids are from the same distribution as the [O I] centroids. A low p-value suggests that the null hypothesis (i.e., the samples are from the same distribution) should be rejected.  (a) From \citet{Banzatti_2019}.
\end{flushleft}
\rtask{table:5}
\end{table*}

In terms of the RV-corrected centroid velocities, eight of the nine components (three for each ensemble group) indicate outflowing emission.  The median values of [O I] are intermediate to those of [1,4] and [0,2] in the SC and BC distributions, but a few km s\textsuperscript{-1} less in the NCs.  Visually, the [O I] component velocities are consistent with zero velocity (i.e., the kinematics indicate Keplerian rotation in a bound disk) more frequently than is the case for the component velocities of fluorescent H\textsubscript{2}.  However, the BC velocities of [O I] span a wide range and do not exhibit a peak.  Following the results of the K-S tests reported in Table \ref{table:5}, the component velocity distributions (grouping SCs, NCs and BCs) are unlikely to sample the same distributions as the stellar RVs, and the RV-corrected H\textsubscript{2} distributions are unlikely to sample the same distribution as [O I].

\begin{figure}
  \centering
  \includegraphics[width=0.87\linewidth]{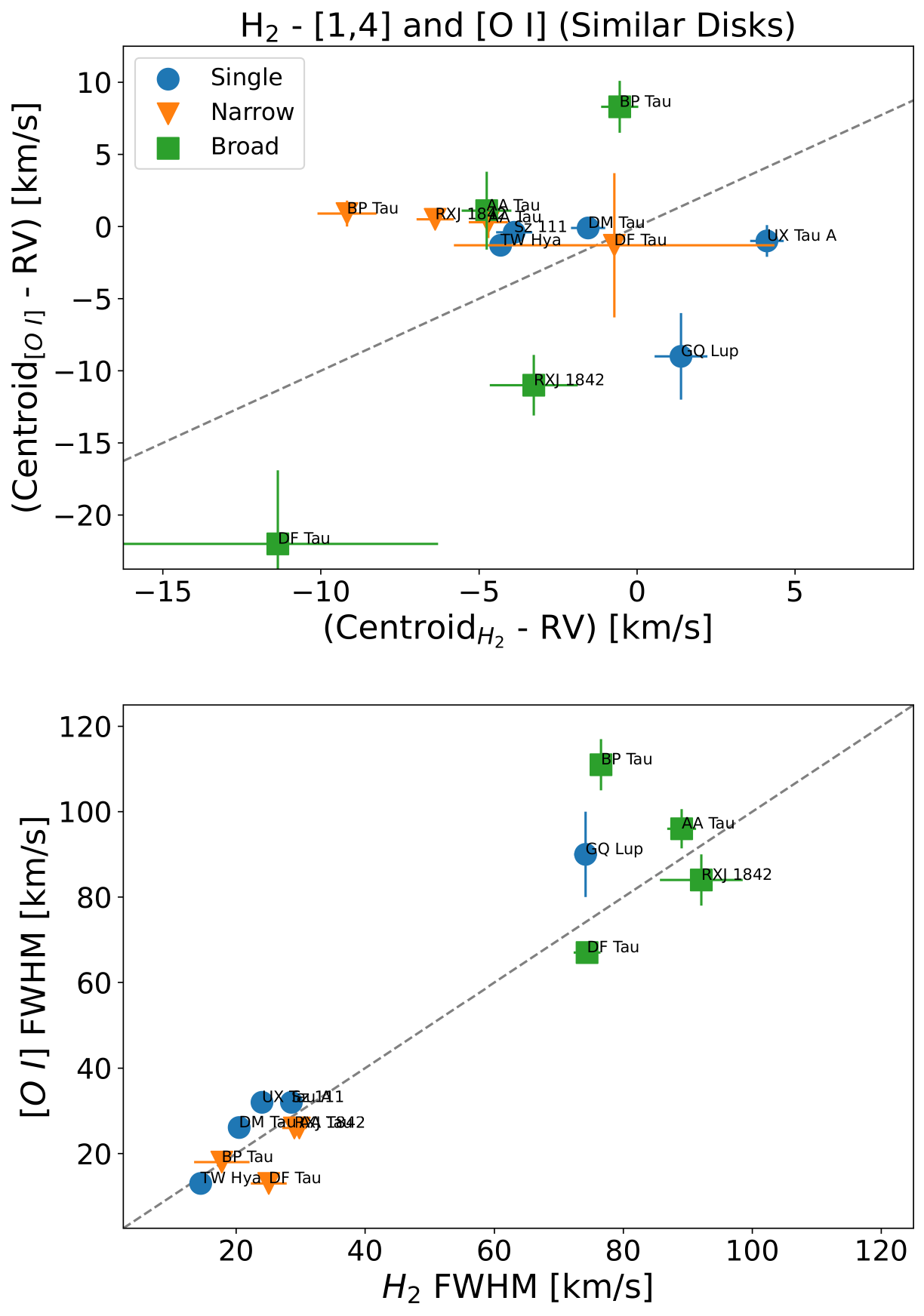}
  \caption{One-to-one comparison of [1,4] and [O I] line kinematics for disks common to both samples, and with identical classification (i.e., SC or NC + BC).  The gray dashed lines represent a hypothetical exact correspondence.  We correct the H\textsubscript{2} centroid velocities for stellar RV using the RVs reported in \citet{Banzatti_2019}.}
  \label{fig:11}
\end{figure}

In terms of FWHM, there is strong overlap between the broad and narrow components of H\textsubscript{2} - [1,4] and [O I].  This translates to similar inferred emitting radii for the two species, consistent within the uncertainties.  There is less overlap between the SC FWHMs, but examination of those disks with identical classification in both [1,4] and [O I] shows a nearly one-to-one correspondence for both SC and NC (bottom panel of Figure \ref{fig:11}).  The correspondence is not as strong for the BCs, but this may reflect difficulties in fit convergence for disks with high-velocity components (see, e.g., Figure 15 of \citealt{Banzatti_2019}).

\section{Discussion} \label{sec:6}

\subsection{Origin of the Fluorescent H\textsubscript{2}}

As we demonstrated in Section \ref{sec:4}, all fluorescent H\textsubscript{2} component groups, with the exception of the single Gaussian components of the [1,4] progression, are systematically blueshifted on the order of a few km s\textsuperscript{-1}.  This suggests a disk-wind origin for some fluorescent H\textsubscript{2} in many CTTS sources, and not solely gas bound to the disk in Keplerian rotation.  While previous studies have discussed outflowing fluorescent H\textsubscript{2} in a subset of T Tauri stars, they were either limited in sample size (e.g., \citealt{Herczeg_2006, Schneider_2013}) or did not systematically consider the possibility of a disk-wind origin (e.g., \citealt{Herczeg_2006, France_2012}).  The small blueshifts involved have more commonly been interpreted as coincident with stellar RVs within the wavelength uncertainty of the observing instrument (see again, e.g., \citealt{Gangi_2023, France_2023}).  Additionally, much of the early discussion was driven by observations and models of H\textsubscript{2} fluorescence in nearby TW Hya, taken to be consistent with an origin in the warm surface layers of the disk (\citealt{Herczeg_2002, Herczeg_2004}).

As we are not able to self-consistently correct component centroid velocities for stellar RV using photospheric lines in the FUV, we must rely on literature values.  These RV estimates can differ on the order of a few km s\textsuperscript{-1} depending on the specific photospheric lines and analyses employed (see, e.g., the discussion in \citealt{Banzatti_2022, Campbell-White_2023b}), adding uncertainty on top of that introduced by the \textit{HST}-COS wavelength solution and the fitting routine.  While this places the velocities measured for a number of disk components within the uncertainties, complicating categorization within the framework of a disk-wind or disk-bound origin, the systematic blueshifts observed in such a large sample cannot be explained by random chance.  This interpretation is supported by K-S tests (see Table \ref{table:5}) which demonstrate a statistically-significant difference between the centroid velocity distributions of fluorescent H\textsubscript{2} and stellar RV.

One peculiarity of our analysis is that [0,2] line centroids are consistently blueshifted an additional few km s\textsuperscript{-1} relative to the [1,4].  This difference persists in the aggregate and on the level of individual disks fit by the same number of Gaussian components.  We explore two possibilities for this difference.

First, we note that all [1,4] emission lines used to perform the co-additions are predominately sampled by the G160M grating of \textit{HST}-COS.  The single line that we use to model [0,2] emission is sampled by the G130M grating.  To test for a potential systematic difference in lines recorded by different gratings, we compare the observed peak velocities of individual emission lines (see, e.g., \citealt{Herczeg_2002} for a list of line possibilities), grouped by grating, in five high-S/N disks of [1,4] and [0,2].  While we note a small systematic difference between the peak velocities measured by our chosen [0,2] line and the 0–6 P(3) ($\lambda$\textsubscript{lab} = 1463.83 {\AA}) line sampled by G160M, we do not see a systematic difference in the peak velocities as measured by the broader groupings in either [0,2] or [1,4].  Therefore, we deem it unlikely that the difference in blueshifted velocities between [1,4] and [0,2] can be plausibly explained by instrumental effects.

A potential explanation for the disparate progression velocities comes by way of spatially-resolved observations of T Tauri with \textit{HST}-STIS (\citealt{Walter_2003}).  The authors observe fluorescent H\textsubscript{2} emission lines in both on-source and extended spectra.  They find that the flux from fluorescent H\textsubscript{2} lines pumped by the red wing of Ly-$\alpha$ (e.g., [0,2] lines) is significantly stronger in on-source spectra, while the flux from fluorescent H\textsubscript{2} lines pumped from near Ly-$\alpha$ line center (e.g., [1,4] lines) contains a substantial extended component (perhaps pumped by local, and not stellar, Ly-$\alpha$).  As \textit{HST}-COS does not spatially resolve the disk from the surrounding circumstellar environment, it may be that velocities of the [1,4] emission lines are reduced with respect to [0,2] due to the increased contribution from extended H\textsubscript{2} gas.  That is, emission originating from near the disk surface may be outflowing at higher velocities than emission observed at extended distances from the disk.

In terms of a MHD or PE wind origin for the blueshifted fluorescent H\textsubscript{2} emission, we refer to the emitting radii reported in Section \ref{sec:4}.  While atomic or ionized PE winds can be launched from $\sim$au radii (e.g., \citealt{Alexander_2014, Pascucci_2023, Sellek_2024}), molecular PE winds (which are cooler) are launched much farther from the star (typically 50 to 100 au for the bulk flow; e.g., \citealt{Gorti_2009}).  They cannot be launched in a thermal PE wind from deep into the gravitational potential well of the star, so a median emitting radius interior to 0.26 au clearly implies a MHD origin for BC disk-wind emission as traced by both fluorescent progressions of H\textsubscript{2}.  This same logic can be applied to the SC disk-wind emission, although emitting radii for a couple of SC disks exceed a few au.  The trend of SC FWHMs increasing with inclination angle is likely a projection effect (i.e., kinematic line broadening is more easily observed in edge-on disks), and not informative vis-à-vis a MHD or PE wind origin.  Similar findings are reported in analyses of [O I] 6300 {\AA} (e.g., \citealt{Simon_2016, Banzatti_2019}), with the caveat that the wind itself could contribute non-trivially to observed broadening (e.g., \citealt{Simon_2016, Shang_2023}).  This is not easily accounted for in point-source spectra, but would affect the values of estimated emitting radii relative to the geometry and magnitude of the wind.

The origin of the NCs is less clear, as inferred emitting radii exceed a few au for many profiles.  In the case of [O I], \citet{Banzatti_2019} argue that correlations between NC and BC line kinematics, together with their correlations to accretion and MHD-launched jet properties, suggest that broad and narrow components are part of the same MHD wind.  We detect correlations between the NC and BC kinematics in both fluorescent H\textsubscript{2} progressions (see Figure \ref{fig:6}), but we do not compare component properties to accretion and fluorescent H\textsubscript{2} emission does not trace jets (associated with high-velocity line centroids) in our sample of CTTSs.  \citet{Banzatti_2019} further argue that the highest blueshifts observed in their sample cluster at disk inclination angles near 35 degrees, which they interpret as a proxy for a 35-degree wind launching angle matching the prediction made by simulations of conical-shell MHD winds (\citealt{Kurosawa_2012}).  We note a slight preference for higher blueshifts at disk inclination angles near 50 degrees (also an intermediate value), but the clustering is tenuous, especially for the single and narrow components, and does not closely match the aforementioned simulations.  Furthermore, synthetic line profiles obtained from X-ray PE and simple MHD models suggest that Gaussian decomposition into narrow and broad components may not allow for clear determination of the respective radial wind-launching regions, and that NC and BC correlations may not necessarily imply a common origin (see \citealt{Weber_2020}).  Therefore, a MHD origin for the ensemble of NC emission remains inconclusive, despite it being favored for those components tracing emission launched closest to their stars.

\subsection{H\textsubscript{2} Mass Loss}

As we argue in favor of a disk-wind origin for a portion of the observed fluorescent H\textsubscript{2}, we also produce simple estimates of H\textsubscript{2} mass loss following the procedure applied to [O I] 6300 {\AA} emission in \citet{Nisini_2024}.  The mass-loss rate can be expressed as 

\begin{equation}
    \dot M\textsubscript{wind} = \mu\, n_{H_{2}}\, \pi\, R^{2}_{w}\, v_{w},
\end{equation}

\vspace{7.5pt}

\noindent where $\mu$ is the molecular mass of H\textsubscript{2}, $n_{H_{2}}$ is the number density of fluorescent H\textsubscript{2}, $R_{w}$ is the radius of the wind region, and $v_{w}$ is the velocity of the wind.  We estimate the number density from typical estimates of the fluorescent H\textsubscript{2} column density (10$^{19}$ cm\textsuperscript{-2}; e.g., \citealt{Herczeg_2004, Hoadley_2017}) and an assumed depth equivalent to the radius of the wind region.  This provides a similar estimate to that obtained in disk atmosphere models (see, e.g., Fig. 5 in \citealt{Pascucci_2023}).  We take the radius of the wind region to be coincident with the component emitting radius.  The wind velocity is as measured by the RV-corrected component velocities that we report in this work.  We do not correct the wind velocity for the disk inclination angle as corrections are modest (e.g., a $\sim$30\% increase for a 50\textdegree\ angle) relative to existing uncertainties.

\begin{table*}
\centering
\textbf{Table 6} \\
\text{H\textsubscript{2} Mass-loss Rates and Mass-loss Efficiencies} \\
\smallskip
 \begin{tabular*}{\textwidth}{c @{\extracolsep{\fill}} cccccc}
 \hline
 \hline
 & \multicolumn{2}{c}{[1,4]} & & \multicolumn{2}{c}{[0,2]} \\
 \cmidrule(lr){2-3} \cmidrule(lr){5-6}
 Name & log($\dot M$\textsubscript{wind})\textsuperscript{(a)} & $\dot M$\textsubscript{wind} / $\dot M$\textsubscript{acc} & ID & log($\dot M$\textsubscript{wind})\textsuperscript{(b)} & $\dot M$\textsubscript{wind} / $\dot M$\textsubscript{acc} & ID \\
 & ($M_{\odot}$ yr\textsuperscript{-1}) & & & ($M_{\odot}$ yr\textsuperscript{-1}) & & \\
 \hline
 AA Tau & -9.25 & 0.17 & NC & -9.42 & 0.12 & NC \\
        & -10.20 & 0.02 & BC & -9.63 & 0.07 & BC \\
 BP Tau & -9.32 & 0.09 & NC & -10.25 & 0.01 & NC \\
        & $\compactcdots$ & $\compactcdots$ & $\compactcdots$ & -10.74 & $<$0.01 & BC \\
 DF Tau & -10.10 & $<$0.01 & BC & -9.81 & 0.01 & BC \\
 ECHA J0843.3-7915 & -10.52 & 0.04 & BC & -10.09 & 0.10 & BC \\
 GM Aur & $\compactcdots$ & $\compactcdots$ & $\compactcdots$ & -9.77 & 0.03 & BC \\
 HN Tau & -9.49 & 0.01 & BC & -9.10 & 0.02 & BC \\
 MY Lup & -8.54 & 0.28 & NC & $\compactcdots$ & $\compactcdots$ & $\compactcdots$ \\
 RX J1556.1-3655 & -9.58 & 0.02 & NC & -8.85 & 0.10 & NC \\
                 & -10.47 & $<$0.01 & BC & -10.04 & 0.01 & BC \\
 RX J1852.3-3700 & -9.12 & 0.38 & NC & -8.31 & 2.46 & NC \\
                 & $\compactcdots$ & $\compactcdots$ & $\compactcdots$ & -9.03 & 0.46 & BC \\
 SY Cha & -9.68 & 0.52 & SC & $\compactcdots$ & $\compactcdots$ & $\compactcdots$ \\
 Sz 71 & -8.86 & 1.52 & NC & $\compactcdots$ & $\compactcdots$ & $\compactcdots$ \\
 Sz 82 & $\compactcdots$ & $\compactcdots$ & $\compactcdots$ & -10.01 & 0.01 & BC \\
 Sz 98 & -10.30 & $<$0.01 & BC & -9.40 & 0.02 & SC \\
 Sz 100 & -9.60 & 2.49 & NC & $\compactcdots$ & $\compactcdots$ & $\compactcdots$ \\
        & -11.13 & 0.07 & BC & $\compactcdots$ & $\compactcdots$ & $\compactcdots$ \\
 Sz 102 & -9.04 & 1.13 & NC & -9.78 & 0.21 & SC \\
        & -10.06 & 0.11 & BC & $\compactcdots$ & $\compactcdots$ & $\compactcdots$ \\
 Sz 103 & -9.86 & 0.28 & NC & -8.97 & 2.14 & NC \\
        & $\compactcdots$ & $\compactcdots$ & $\compactcdots$ & -10.21 & 0.12 & BC \\
 Sz 114 & -11.70 & $<$0.01 & SC & -10.33 & 0.08 & NC \\
        & $\compactcdots$ & $\compactcdots$ & $\compactcdots$ & -11.12 & 0.01 & BC \\
 Sz 129 & -9.75 & 0.04 & NC & -9.71 & 0.04 & SC \\
 TW Hya & -10.97 & 0.05 & SC & $\compactcdots$ & $\compactcdots$ & $\compactcdots$ \\
 V4046 Sgr & $\compactcdots$ & $\compactcdots$ & $\compactcdots$ & -9.89 & 0.01 & SC \\
 \hline
 \end{tabular*}
\begin{flushleft}
\textbf{Notes.} (a) H\textsubscript{2} mass-loss rate estimated from the [1,4] progression given assumption of $\sim$2500 $\pm$ 500 K UV-H\textsubscript{2} gas occupying tail of Maxwell–Boltzmann distribution of warm (500 K) H\textsubscript{2} gas.  (b) As before, but estimated from the parameters of the [0,2] model fits.  Typical uncertainties are an order of magnitude for $\dot M$\textsubscript{wind} estimates.
\end{flushleft}
\rtask{table:6}
\end{table*}

\begin{figure}
  \centering
  \includegraphics[width=0.87\linewidth]{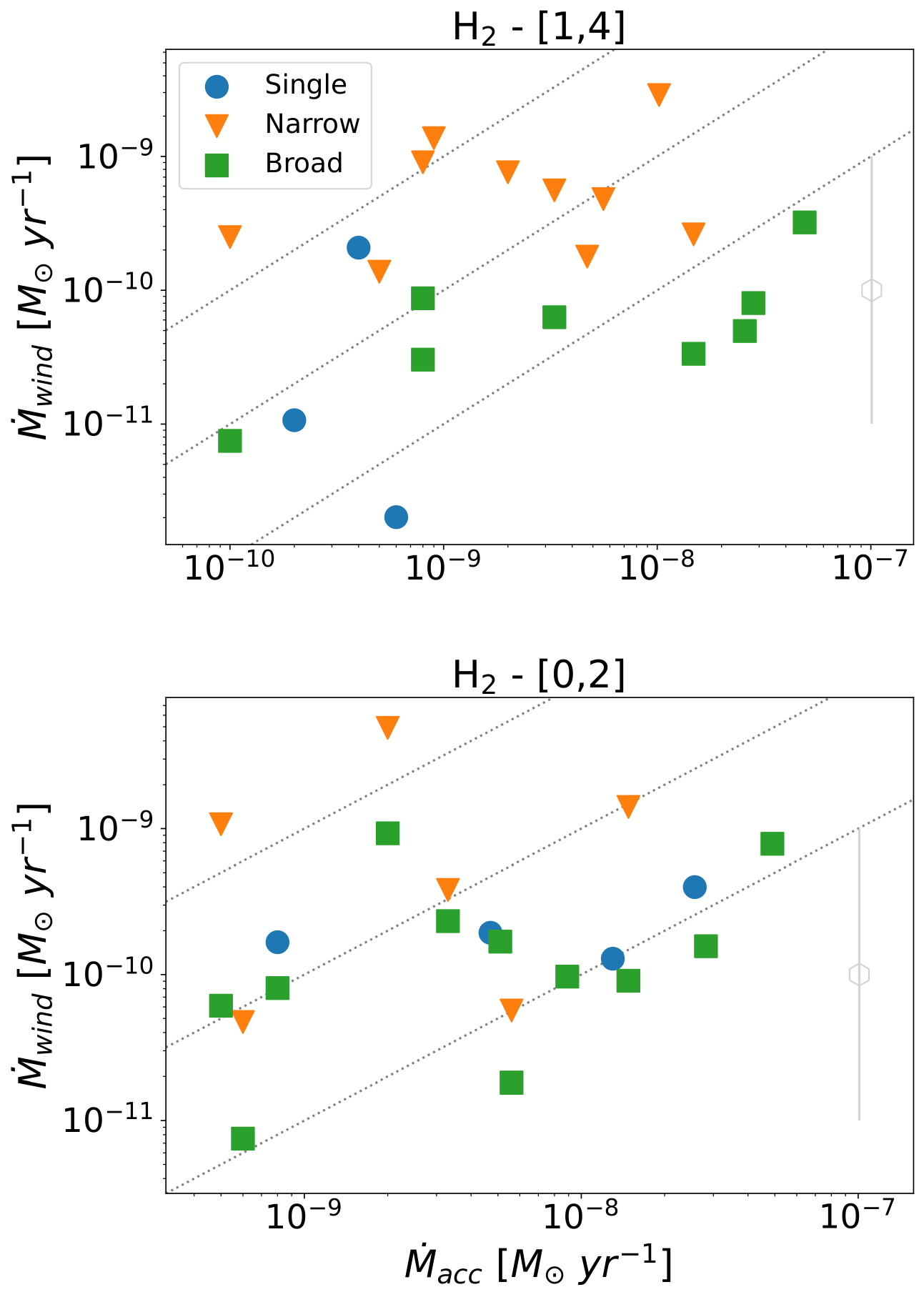}
  \caption{Correlation between H\textsubscript{2} mass-loss rates and stellar mass accretion.  Uncertainties on individual mass-loss estimates span at least an order of magnitude, which we represent with light gray error bars on the right side of each plot.  We include lines of constant mass-loss efficiency (0.01, 0.1 and 1) as dotted gray lines.}
  \label{fig:12}
\end{figure}

As fluorescent H\textsubscript{2} emission traces the hot ($\sim$2500 K) molecular gas at the disk surface (e.g., \citealt{Herczeg_2004}), it does not account for the warm ($\sim$500 K) molecular component which likely represents a much larger fraction of the molecular mass near the disk surface (see, e.g., \citealt{Carr_2008} and \citealt{Xie_2023} for examples of warm molecular emission in the inner disk), and which appears to be outflowing in recent JWST spectro-imaging of infrared H\textsubscript{2} lines in small samples of edge-on T Tauri disks (e.g., \citealt{Arulanantham_2024, Pascucci_2024}).  To account for this potential discrepancy, and thus produce better estimates of total H\textsubscript{2} mass loss, we model the fluorescent H\textsubscript{2} population as occupying the 2000 to 3000 K tail of a Maxwell–Boltzmann distribution of 500 K H\textsubscript{2} gas.  We divide our mass-loss estimates by the fractional area of the distribution occupied by the hot gas, increasing their calculated values by about an order of magnitude.  We report H\textsubscript{2} mass-loss rates and efficiencies, calculated separately for each disk component with measured wind velocities in excess of the wavelength uncertainty of \textit{HST}-COS ($\sim$5 km s\textsuperscript{-1}), in Table \ref{table:6}.  We directly compare H\textsubscript{2} mass-loss rates to literature values of stellar accretion rates in Figure \ref{fig:12}.

H\textsubscript{2} mass-loss rates, as estimated by both [1,4] and [0,2] model components, span a range of magnitudes from approximately 10$^{-9}$ to 10$^{-11}$ $M_{\odot}$ yr\textsuperscript{-1}, with no clear correlation to the magnitude of stellar mass accretion.  Mass-loss efficiencies ($\dot M$\textsubscript{wind}/$\dot M$\textsubscript{acc}) are largely clustered between 0.01 and 1.  The majority of components with mass-loss efficiencies larger than 1 pertain to dust-evolved disks ($n$\textsubscript{13-31} $>$ 0).  This may point to a scenario where disk winds persist in the gaseous disk (perhaps launched from progressively larger radii) as the dust disk evolves and stellar accretion rates decline.

Overall, the calculated mass-loss and mass-loss efficiency rates are consistent with those calculated via other disk wind tracers (see, e.g., \citealt{Fang_2018, Nisini_2024}).  They also agree, within an order of magnitude, with global simulations of MHD wind mass-loss rates in the inner disk (Fig. 6 of \citealt{Pascucci_2023}).  This may support a scenario where both the broad and narrow components of fluorescent H\textsubscript{2} trace a MHD wind.

One important caveat of this discussion is that uncertainties on the parameters we use to calculate mass loss are large.  In addition to the uncertainties on the emitting radii and wind velocities estimated by the fitting routine, we assume a generic H\textsubscript{2} column density which does not account for a potential density gradient in the inner disk.  We also assume that the wind is launched from a region encompassing the entire disk interior to the emitting radius, as well as the region's vertical extent.  Fundamentally, any estimates of wind mass loss via point-source spectra are complicated by a lack of spatial information, with limited knowledge of wind geometry and physical gradients in the disk and wind, implying uncertainties of at least an order of magnitude (see \citealt{Nisini_2024} for a more complete discussion).

\subsection{True Magnitude of Blueshifted Emission}

In associating emission from individual Gaussian components to single peak centroid velocities, we make the tacit assumption that the emission contained within each component is outflowing (or not) in unison.  In reality, individual components (especially components in SC sources) may be a blend of both outflowing and disk-bound material.  While we cannot readily disentangle these varied contributions in point-source spectra, we can perform a simple test to estimate the magnitude of blueshifted emission needed to recreate SC line profiles of different peak velocities.

\begin{figure}
  \centering
  \includegraphics[width=0.95\linewidth]{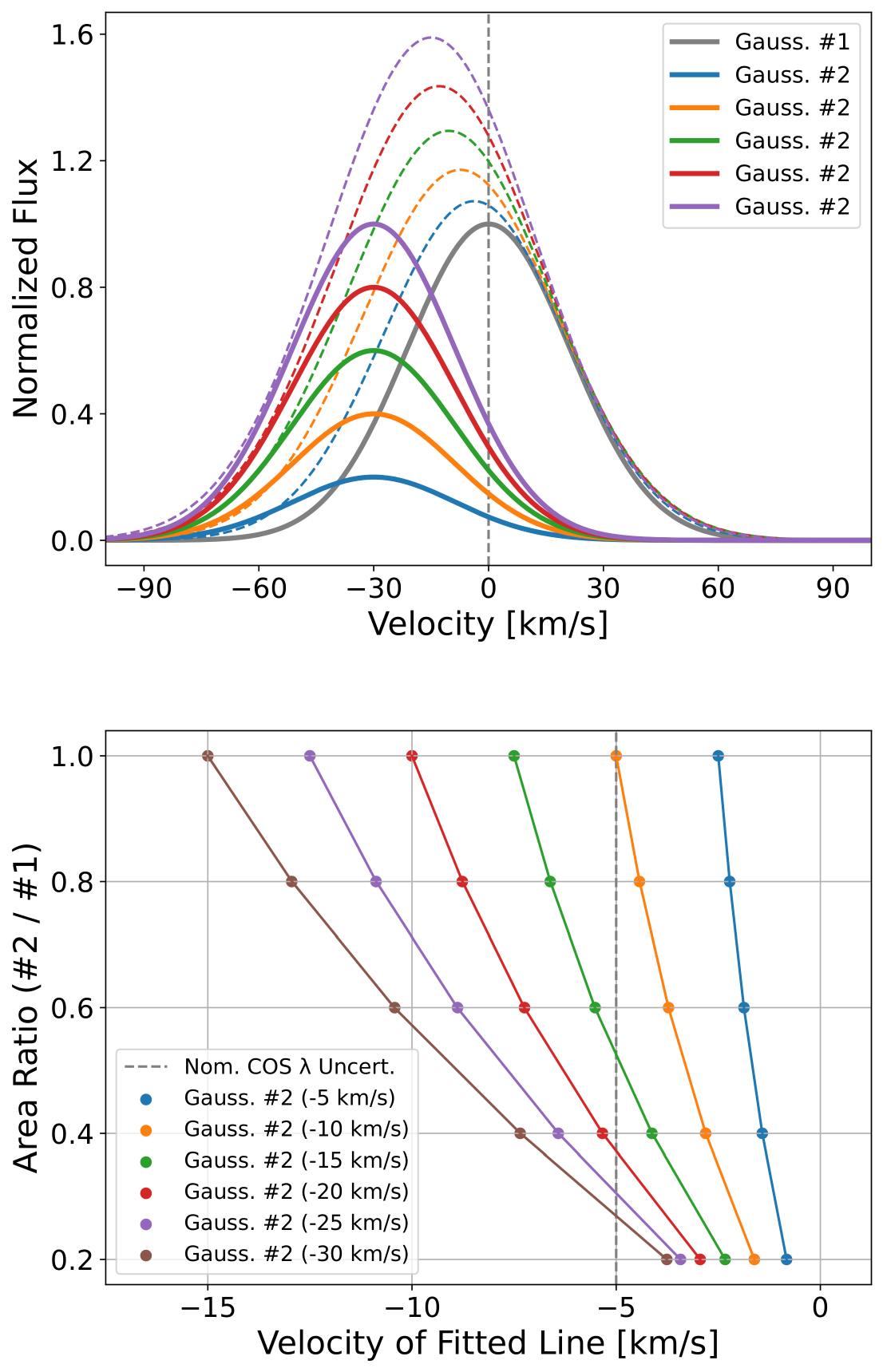}
  \caption{Procedure for estimating the velocity and quantity of blueshifted emission needed to reproduce different SC line profiles.  In the top panel, the 0 km s\textsuperscript{-1} Gaussian component is illustrated by a solid gray line and the -30 km s\textsuperscript{-1} Gaussian components of varying amplitude by solid colored lines.  The single Gaussian fits to their superpositions (not shown) are marked by the corresponding dashed colored lines.  In the bottom panel, the degree of blueshifted emission is compared to its effect on the velocity of the fitted line.  The nominal wavelength uncertainty of \textit{HST}-COS is marked by the dashed gray line.}
  \label{fig:13}
\end{figure}

We simulate SC line profiles by fitting a single Gaussian to superpositions of two Gaussian components, one fixed at 0 km s\textsuperscript{-1} (i.e., bound to the disk) and the other blueshifted (i.e., outflowing) between -5 and -30 km s\textsuperscript{-1} (following the values measured in our sample).  We set the FWHM of both components to 50 km s\textsuperscript{-1}, but vary the amplitude of the blueshifted component between 20 and 100\% of the amplitude of the fixed component.  An illustration of this procedure is displayed in the top panel of Figure \ref{fig:13}, wherein the single Gaussian fits to the superposed components are shown as dashed lines.

To quantify the amount of blueshifted emission needed to produce different single Gaussian fits, we plot the velocities of the fitted lines versus the area ratios of the blueshifted and fixed components in the bottom panel of Figure \ref{fig:13}.  To reproduce a SC profile peaked at -5 km s\textsuperscript{-1} (the approximate wavelength uncertainty of \textit{HST}-COS), one could surmise that half of the emission is bound to the disk, while the other half is outflowing at -10 km s\textsuperscript{-1}.  A faster outflow would require less blueshifted emission to reach the same SC peak velocity, and vice versa.

While these results are perhaps not surprising, they may support the existence of a small blueshifted wind component even in those Gaussian components identified as consistent with stellar RV.  That is, there can be some level of blueshifted emission below which there is minimal measured effect on the location or shape of the observed line.  If so, we may be underestimating the frequency of fluorescent H\textsubscript{2} outflows in CTTSs.

\subsection{Cartoon View of Fluorescent H\textsubscript{2} and [O I] 6300 {\AA} Emission in CTTSs}

\begin{figure*}
  \centering
  \includegraphics[width=\textwidth]{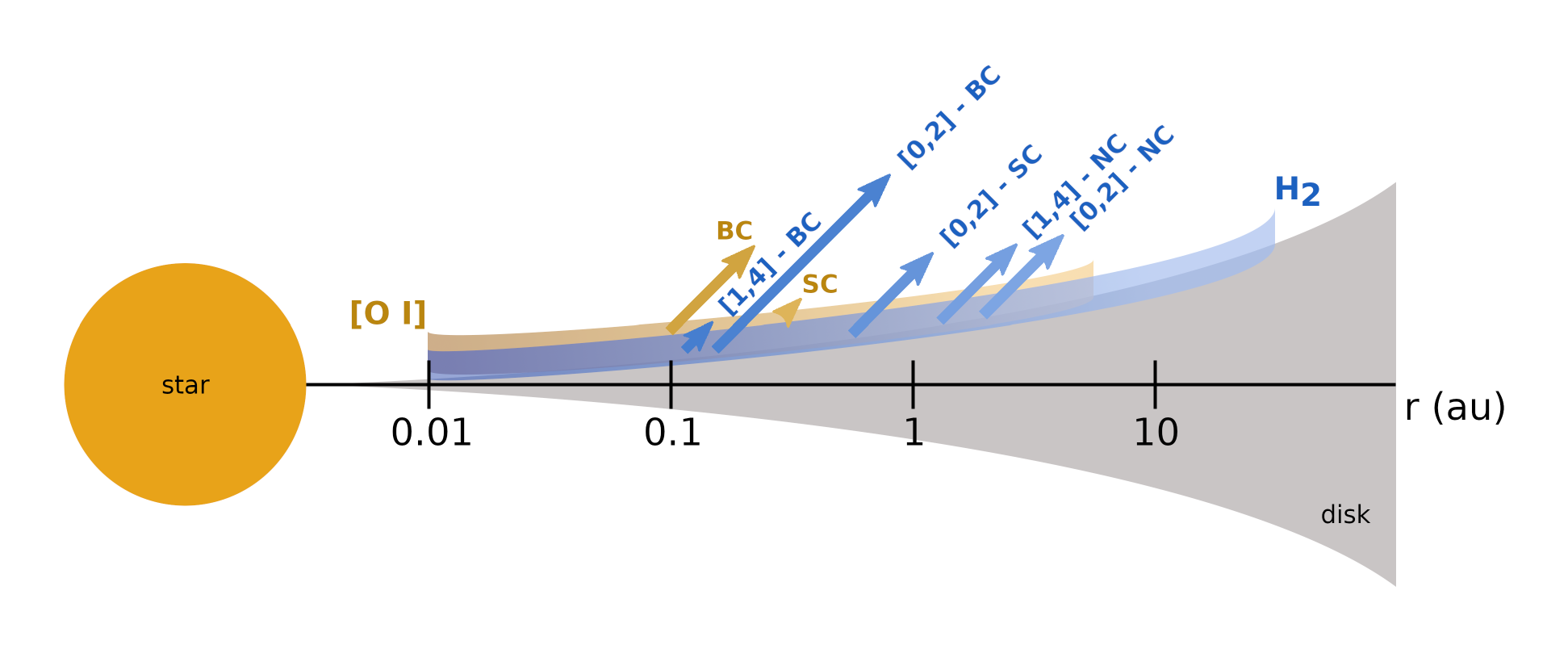}
  \caption{Schematic of fluorescent H\textsubscript{2} (in blue) and [O I] 6300 {\AA} (in gold) emission in a CTTS.  The magnitudes of median component wind velocities are represented by the lengths of the arrows.  The arrows are placed according to the medians of the component emitting radii.  We exclude the SCs of H\textsubscript{2} - [1,4] and the NCs of [O I] because their median velocities are near zero.  We illustrate the [O I] emission layer above the H\textsubscript{2} layer because it traces a hotter gas.  The horizontal axis is in log scale.  Adapted from \citet{Gangi_2023}.  \\}
  \label{fig:14}
\end{figure*}

Following the results of our analysis, and comparison with the results of \citet{Banzatti_2019}, we present a cartoon view of fluorescent H\textsubscript{2} and [O I] 6300 {\AA} emission in CTTSs (Figure \ref{fig:14}).  From the component FWHMs and inferred emitting radii, H\textsubscript{2} and [O I] emission spatially overlap in the innermost disk ($\sim$0.1 au) out to $\sim$10 au.  In many disks, atomic and molecular winds are launched primarily between 0.1 and 3 au of the central star, the magnitudes of which are consistent on the order of a few km s\textsuperscript{-1}.  These common spatial and kinematic properties are indicative of a multi-component disk wind, in which atomic and molecular gas are entrained in the same outflow in some sources.  As both fluorescent H\textsubscript{2} and [O I] 6300 {\AA} trace hot gas (e.g., \citealt{Herczeg_2004, Natta_2014}), we show that they are launched from near the surface of the disk atmosphere.

\section{Summary} \label{sec:7}

We have analyzed fluorescent H\textsubscript{2} emission, traced by two progressions in the FUV, for evidence of a disk-wind origin in a sample of 36 high-S/N CTTSs, predominately observed in the framework of \textit{HST}'s ULLYSES program.  We used a MCMC fitting routine to classify co-added profiles in both progressions on the basis of the number of model Gaussian components needed to best fit the data.  Following the disk-wind literature, we labelled the Gaussian component of one-component disks as SC, and the narrow and broad Gaussian components of two-component disks as NC and BC, respectively.  We measured component kinematic line properties (centroid velocity and FWHM), inferred emitting radii, and compared our measurements to those obtained via atomic tracers of low-velocity disk winds.  Our main findings are as follows: 

\begin{enumerate}
    \item The [1,4] fluorescent progression of H\textsubscript{2}, used to make the cut from the original list of 91 CTTSs, is detected at high S/N in 36 sources.  The [0,2] progression is detected at high S/N in the majority (64\%) of those sources.  Co-added line profiles of both progressions are best fit by two Gaussian components and a linear continuum in two-thirds of high-S/N sources.
    \item The [1,4] and [0,2] fluorescent progressions of H\textsubscript{2} trace a blueshifted low-velocity disk wind outflowing faster than the wavelength uncertainty of \textit{HST}-COS ($\sim$5 km s\textsuperscript{-1}) in  at least one component of 50 and 70\% of sources, respectively.  In the ensemble of all high-S/N disks, the magnitude of this wind is on the order of a few km s\textsuperscript{-1}, with larger negative velocities observed for gas traced by [0,2].  The origin of this magnitude difference is unclear, but may be due to different spatial extents of the two progressions.
    \item These low-velocity winds are launched primarily between 0.1 and 3 au of the central star.  The BCs very likely trace a MHD wind launched interior to 0.5 au.  A MHD or PE origin for the NCs is less clear; however, NC and BC line kinematics are correlated and emitting radii ($<$1 or 2 au) for several NCs favor a magnetic origin.
    \item We do not see strong correlations between line kinematics, disk inclination and the $n$\textsubscript{13-31} infrared index.  As such, we cannot confidently speak of a preferred launching angle for H\textsubscript{2} disk winds, or how they may co-evolve with the dusty disk.
    \item Fluorescent H\textsubscript{2} and [O I] 6300 {\AA} line kinematics are broadly similar, although [O I] centroids are more frequently consistent with zero velocity.  They both trace emission over a range of radii ($\sim$0.1 to 10 au) in the inner disk.
    \item Fluorescent H\textsubscript{2} emission can be used to estimate molecular mass-loss rates from the inner disk, and those estimates (10$^{-9}$ to 10$^{-11}$ $M_{\odot}$ yr\textsuperscript{-1}) are similar to estimates made from atomic wind tracers and global simulations.  However, uncertainties are large and spatial information is needed to better evaluate how molecular wind mass loss affects disk dispersal.
    \item The true degree of blueshifted emission contained within Gaussian component fits to point-source spectra is hard to quantify, but SC profiles can be reproduced by different combinations of disk-bound and disk-wind material.  This suggests that profiles with component centroid velocities near the stellar RV may be hiding a small amount of outflowing emission.
\end{enumerate}

Overall, our analysis paints a picture of a multi-component disk wind, where atomic and molecular gas are launched from similar radii, and at similar velocity, in many CTTS sources.  It highlights the potential of fluorescent H\textsubscript{2} as an additional tracer of disk outflows.  As it does not usually trace MHD jets in CTTSs, the low-velocity components are free of blending with a high-velocity component, which may simplify the analysis of line parameters relative to atomic wind tracers.  In the era of JWST, we may be able to combine fluorescent H\textsubscript{2} line properties with spatially-resolved, mid-infrared H\textsubscript{2} spectra (e.g., \citealt{Arulanantham_2024}) to better understand the role that molecular winds play in the evolution and dispersal of protoplanetary disks.

\section*{Acknowledgements}

This work was carried out as part of the Outflows and Disks around Young Stars: Synergies for the Exploration of Ullyses Spectra (ODYSSEUS) program, supported by \textit{HST} archival grant number 16129-025 to the University of Colorado Boulder.  B.N. acknowledges support from Large Grant INAF 2022 “YSOs Outflows, Disks and Accretion: towards a global framework for the evolution of planet forming systems (YODA)” and PRIN-MUR 2022 20228JPA3A “The path to star and planet formation in the JWST era (PATH).”  R.A. acknowledges funding from the Science \& Technology Facilities Council (STFC) through Consolidated Grant ST/W000857/1.  J.C.-W. acknowledges funding by the European Union under the Horizon Europe Research \& Innovation Programme 101039452 (WANDA).  Views and opinions expressed are, however, those of the author(s) only and do not necessarily reflect those of the European Union or the European Research Council.  Neither the European Union nor the granting authority can be held responsible for them.  Z.G. is supported by the ANID FONDECYT Postdoctoral program No. 3220029 and ANID, Millennium Science Initiative, AIM23-0001.

Based on observations obtained with the NASA/ESA \textit{Hubble Space Telescope}, retrieved from the Mikulski Archive for Space Telescopes (MAST) at the Space Telescope Science Institute (STScI).  The specific ULLYSES observations analyzed can be accessed via \dataset[https://doi.org/10.17909/t9-jzeh-xy14]{https://doi.org/10.17909/t9-jzeh-xy14}.  STScI is operated by the Association of Universities for Research in Astronomy, Inc. under NASA contract NAS 5-26555.

\section*{Appendix A} \label{appendix:A}

Figure \ref{fig:15.1} shows the co-added line profiles and Gaussian model fits for every disk and progression which met the S/N threshold, as described in Section \ref{sec:3}.  The [1,7] and [3,16] fit parameters, together with their calculated emitting radii, are reported in Table \ref{table:7}.

\begin{figure*}
  \centering
  \includegraphics[height=1.15\textwidth]{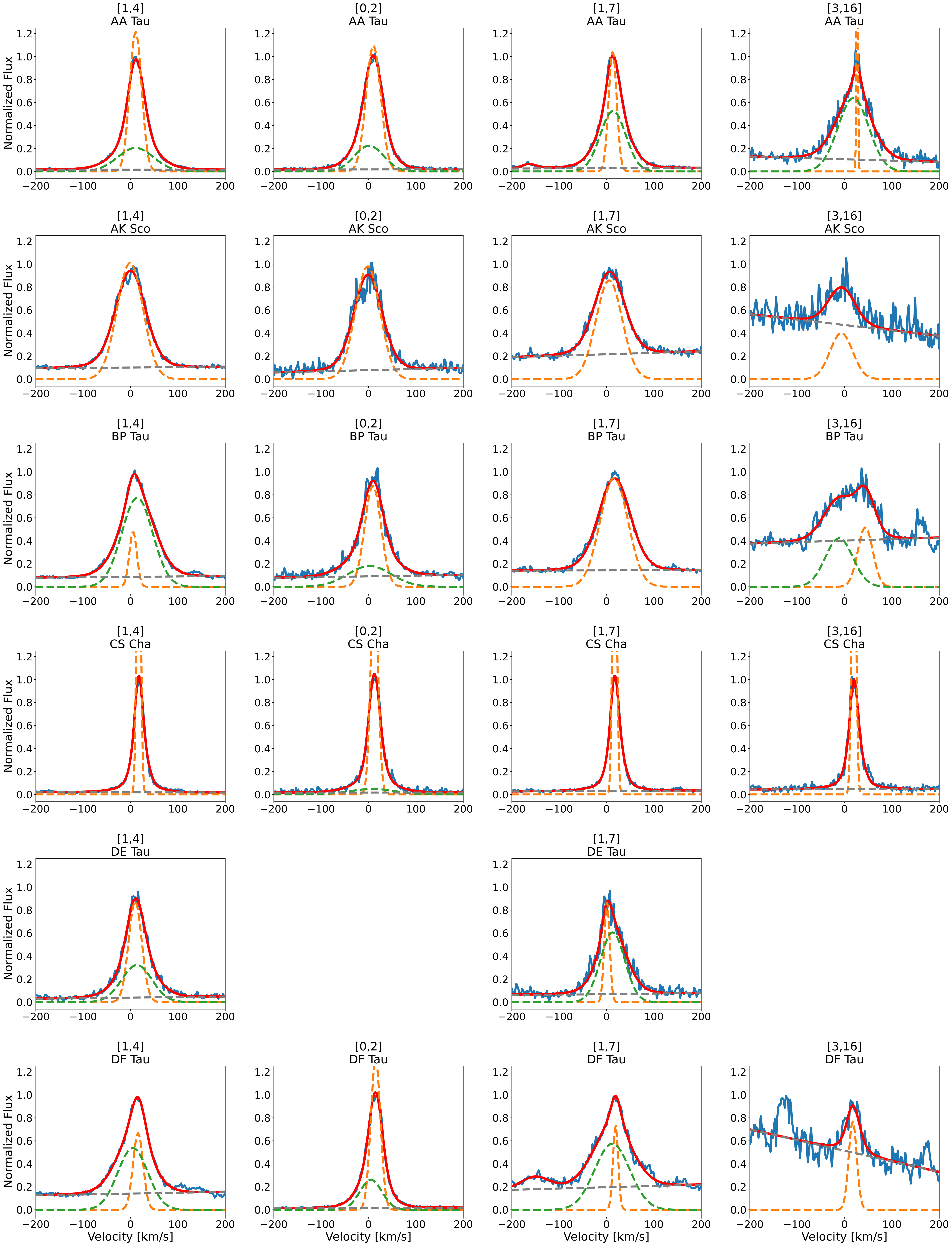}
  \caption{Model fits for all disks which met the S/N threshold in each progression.  Line profiles are fit with either one or two Gaussian components based on the criteria (e.g., $\Delta$BIC) discussed in the text.  The co-added profiles are in blue and the best-fit models in red.  Single and narrow Gaussian components are marked with orange dashes, broad Gaussian components with green dashes, and the linear continuum with gray dashes.  The sum of the broad and narrow Gaussian components exceeds the height of the best-fit model because the best-fit model is convolved with the \textit{HST}-COS LSF (not shown).  Velocities are not corrected for the stellar reference frame.}
  \label{fig:15.1}
\end{figure*}

\begin{figure*}
  \centering
  \addtocounter{figure}{-1}
  \includegraphics[height=1.15\textwidth]{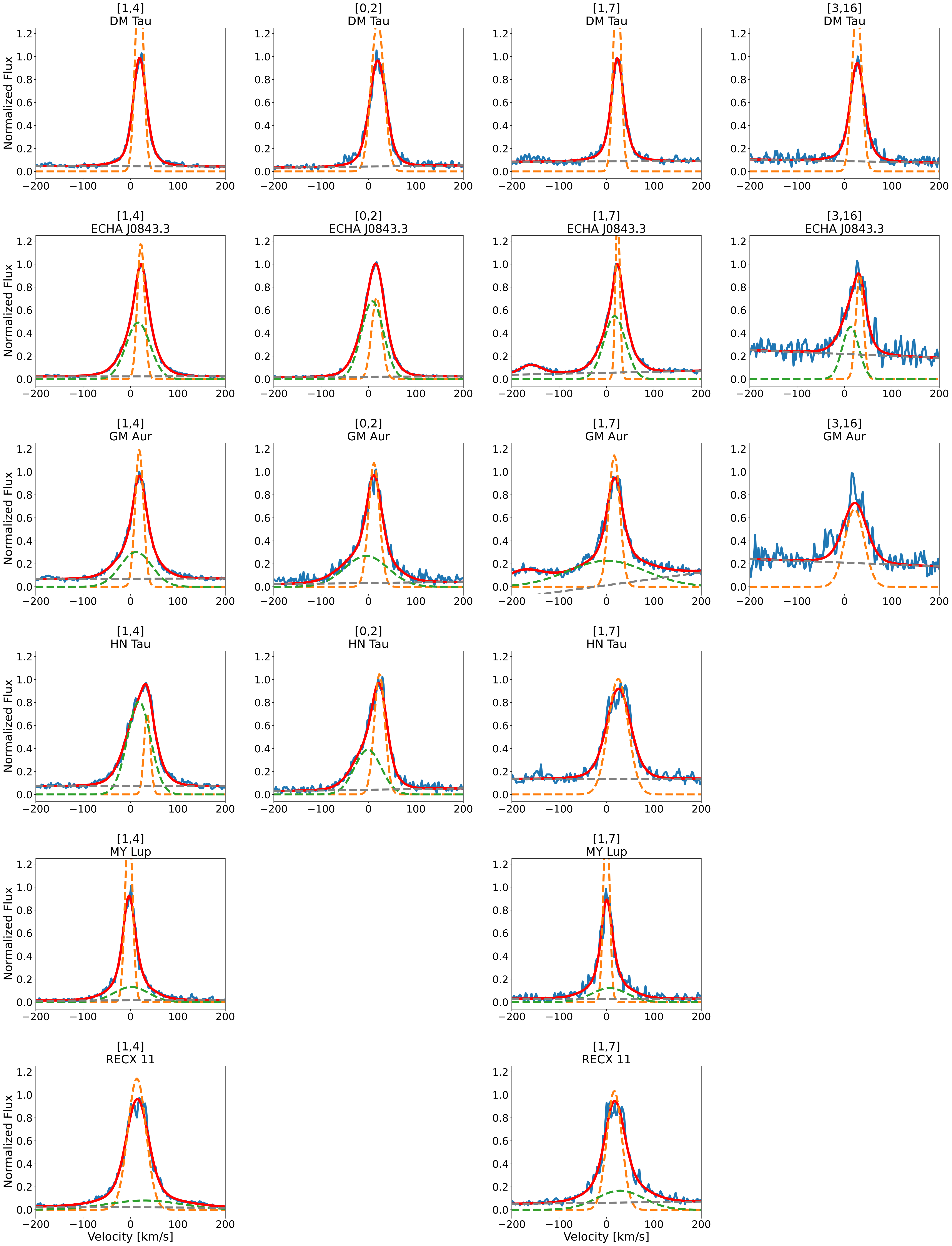}
  \caption{(Continued)}
  \label{fig:15.2}
\end{figure*}

\begin{figure*}
  \centering
  \addtocounter{figure}{-1}
  \includegraphics[height=1.15\textwidth]{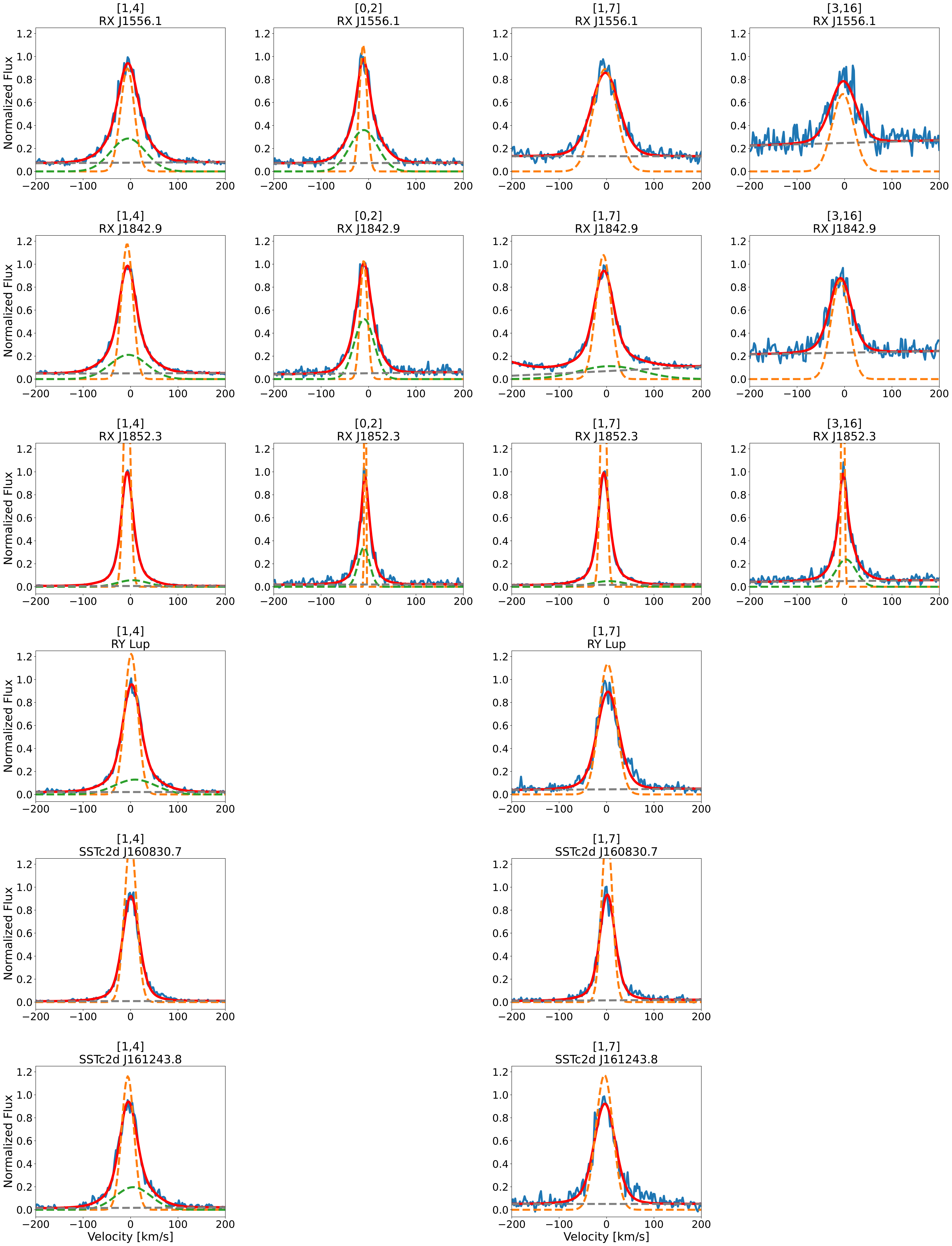}
  \caption{(Continued)}
  \label{fig:15.3}
\end{figure*}

\begin{figure*}
  \centering
  \addtocounter{figure}{-1}
  \includegraphics[height=1.15\textwidth]{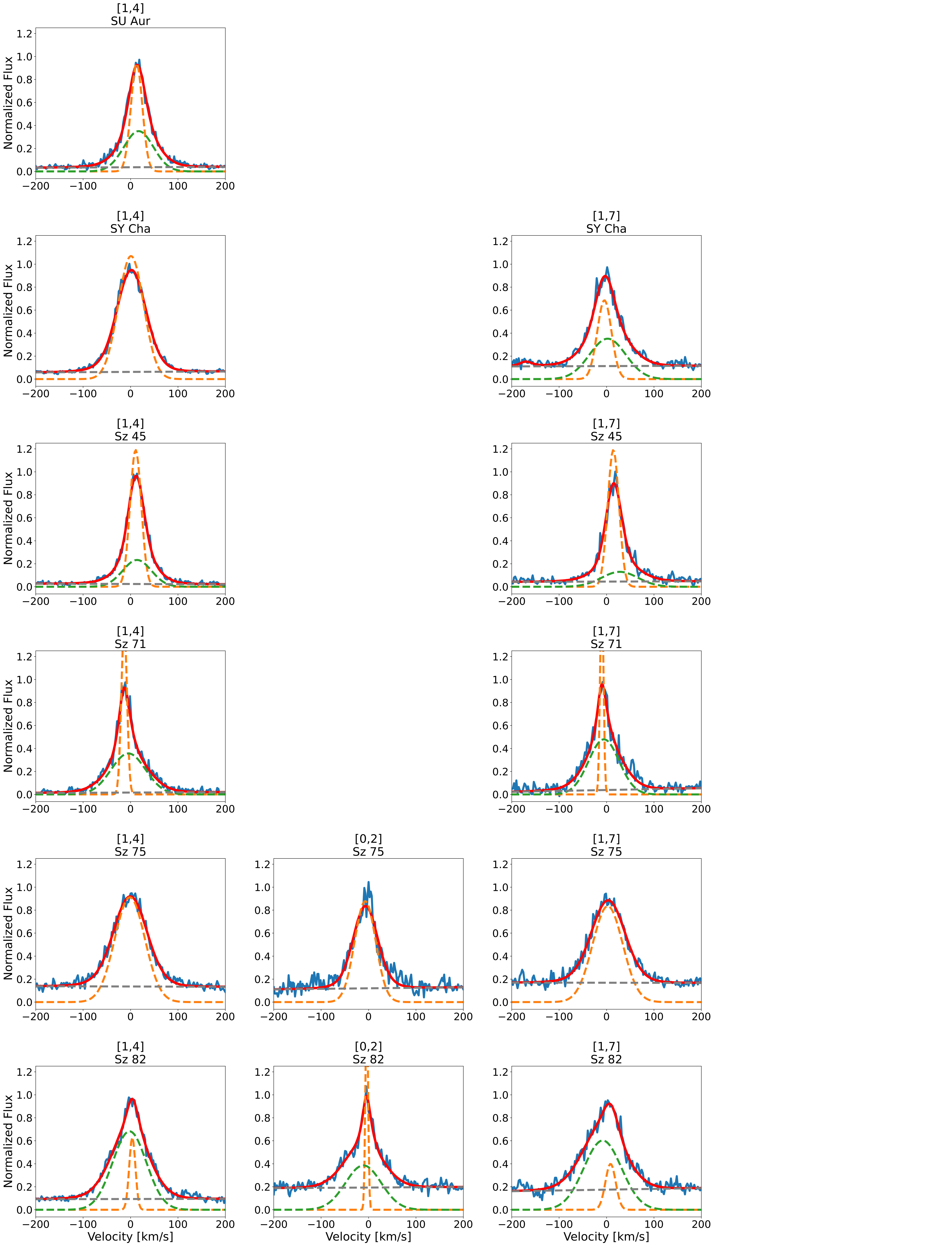}
  \caption{(Continued)}
  \label{fig:15.4}
\end{figure*}

\begin{figure*}
  \centering
  \addtocounter{figure}{-1}
  \includegraphics[height=1.15\textwidth]{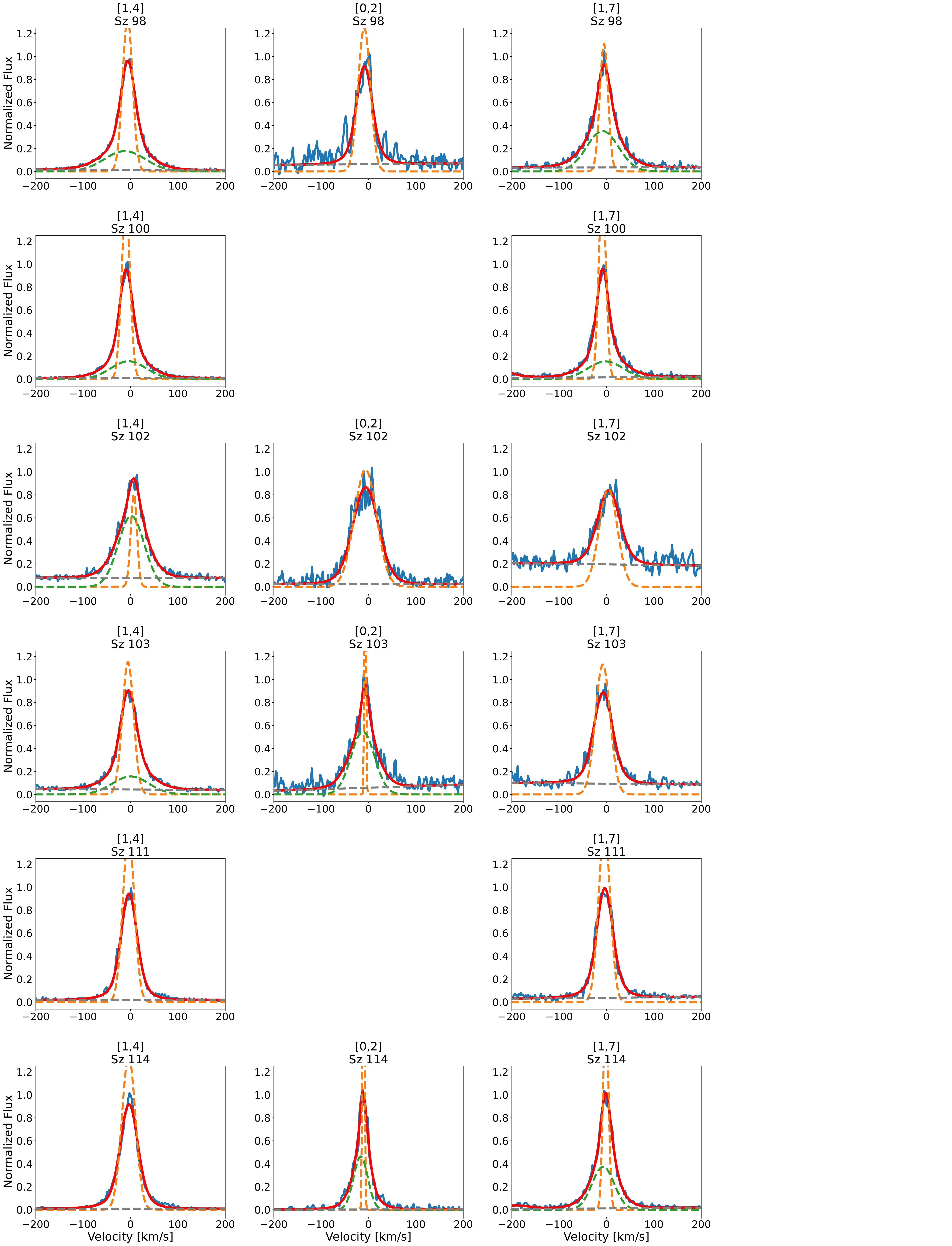}
  \caption{(Continued)}
  \label{fig:15.5}
\end{figure*}

\begin{figure*}
  \centering
  \addtocounter{figure}{-1}
  \includegraphics[height=1.15\textwidth]{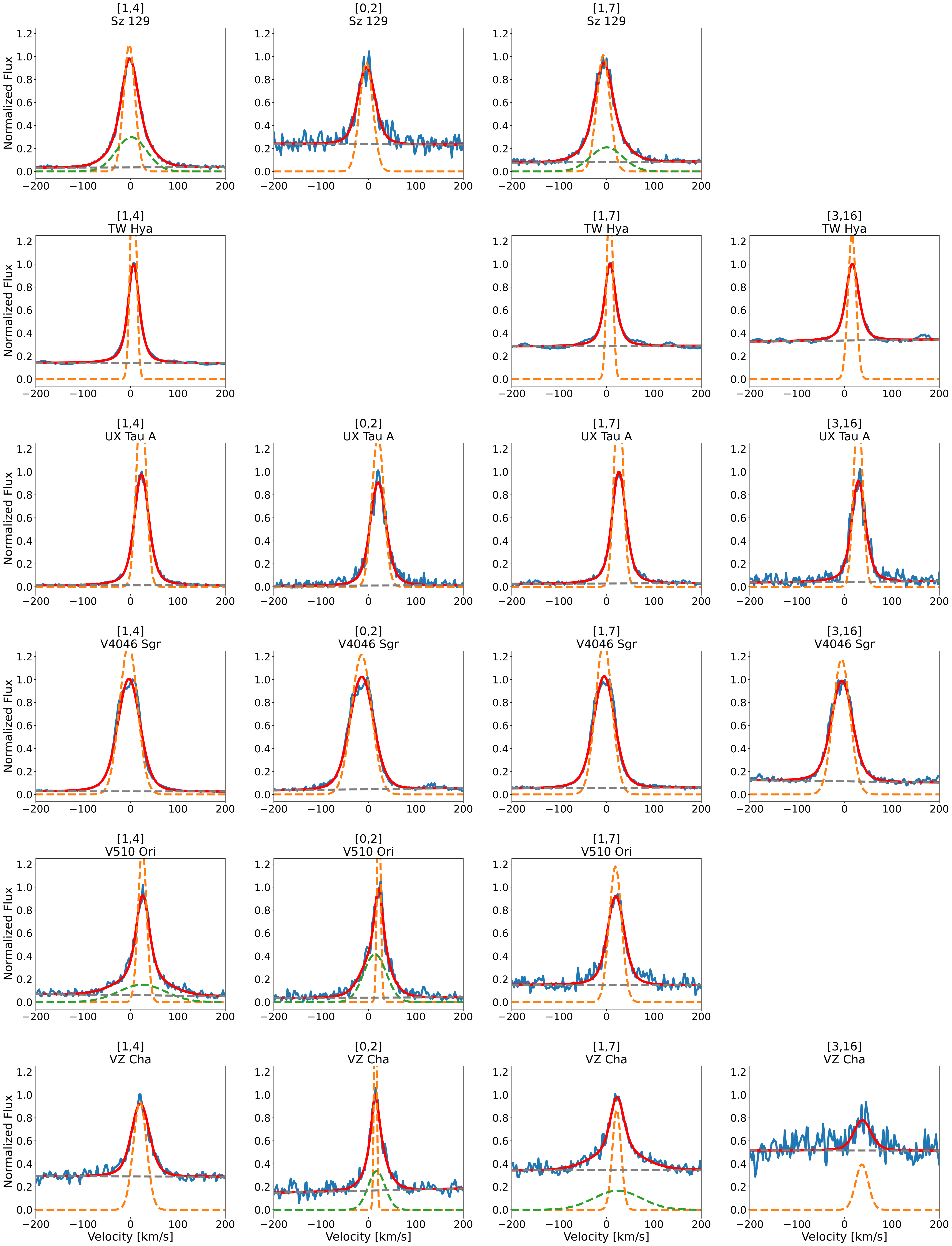}
  \caption{(Continued)}
  \label{fig:15.6}
\end{figure*}

\begin{table*}
\centering
\textbf{Table 7} \\
\text{Properties of the H\textsubscript{2} Emission} \\
\smallskip
 \begin{tabular*}{\textwidth}{c @{\extracolsep{\fill}} cccccccc}
 \hline
 \hline
 & \multicolumn{3}{c}{[1,7]} & & \multicolumn{3}{c}{[3,16]} \\
 \cmidrule(lr){2-4} \cmidrule(lr){6-8}
 Name & Centroid - RV\textsuperscript{(a)} & FWHM\textsuperscript{(b)} & $R$(H\textsubscript{2})\textsuperscript{(c)} & ID\textsuperscript{(d)} & Centroid - RV & FWHM & $R$(H\textsubscript{2}) & ID \\
 & (km s\textsuperscript{-1}) & (km s\textsuperscript{-1}) & (au) & & (km s\textsuperscript{-1}) & (km s\textsuperscript{-1}) & (au) & \\
 \hline
 AA Tau & $-5.16^{+0.43}_{-0.46}$ & $19.45^{+2.26}_{-2.36}$ & $5.51^{+1.28}_{-1.34}$ & NC 
        & $8.10^{+1.80}_{-1.73}$ & $4.37^{+7.42}_{-2.69}$ & $109.22^{+370.90}_{-109.22}$ & NC \\
        & $-4.35^{+0.47}_{-0.48}$ & $62.11^{+2.20}_{-1.97}$ & $0.54^{+0.04}_{-0.03}$ & BC 
        & $0.45^{+1.37}_{-1.45}$ & $77.53^{+3.51}_{-3.62}$ & $0.35^{+0.03}_{-0.03}$ & BC \\
 AK Sco & $6.84^{+2.66}_{-2.66}$ & $64.89^{+1.52}_{-1.50}$ & $0.98^{+0.05}_{-0.05}$ & SC 
        & $-5.04^{+3.82}_{-3.81}$ & $54.78^{+8.78}_{-8.23}$ & $1.37^{+0.44}_{-0.41}$ & SC \\
 BP Tau & $-0.28^{+0.38}_{-0.38}$ & $69.45^{+0.87}_{-0.91}$ & $0.09^{+0.00}_{-0.00}$ & SC 
        & $27.14^{+3.78}_{-5.24}$ & $37.26^{+8.81}_{-8.06}$ & $0.31^{+0.15}_{-0.13}$ & NC \\
        & $\compactcdots$ & $\compactcdots$ & $\compactcdots$ & $\compactcdots$ 
        & $-27.05^{+6.00}_{-8.31}$ & $70.26^{+8.65}_{-10.71}$ & $0.09^{+0.02}_{-0.03}$ & BC \\
 CS Cha & $1.76^{+0.43}_{-0.43}$ & $9.52^{+1.06}_{-1.13}$ & $14.89^{+3.32}_{-3.53}$ & SC 
        & $4.22^{+0.54}_{-0.54}$ & $9.80^{+1.74}_{-1.94}$ & $14.05^{+4.99}_{-5.56}$ & SC \\
 DE Tau & $-10.21^{+4.78}_{-4.55}$ & $37.15^{+15.32}_{-26.67}$ & $0.21^{+0.17}_{-0.21}$ & NC 
        & $\compactcdots$ & $\compactcdots$ & $\compactcdots$ & $\compactcdots$ \\
        & $-1.45^{+35.31}_{-7.53}$ & $61.17^{+24.49}_{-32.61}$ & $0.08^{+0.06}_{-0.08}$ & BC 
        & $\compactcdots$ & $\compactcdots$ & $\compactcdots$ & $\compactcdots$ \\
 DF Tau & $3.65^{+1.64}_{-3.80}$ & $13.57^{+60.18}_{-6.24}$ & $7.46^{+66.15}_{-6.86}$ & NC 
        & $2.47^{+1.67}_{-1.62}$ & $18.76^{+5.19}_{-5.87}$ & $3.90^{+2.16}_{-2.44}$ & SC \\
        & $-1.87^{+4.87}_{-1.27}$ & $83.88^{+5.53}_{-70.03}$ & $0.20^{+0.03}_{-0.20}$ & BC 
        & $\compactcdots$ & $\compactcdots$ & $\compactcdots$ & $\compactcdots$ \\
 DM Tau & $3.64^{+0.35}_{-0.36}$ & $18.40^{+1.33}_{-1.32}$ & $1.72^{+0.25}_{-0.25}$ & SC 
        & $7.29^{+0.60}_{-0.62}$ & $23.46^{+1.99}_{-1.95}$ & $1.06^{+0.18}_{-0.18}$ & SC \\
 ECHA J0843.3 & $1.14^{+0.56}_{-0.54}$ & $10.33^{+3.32}_{-3.18}$ & $4.99^{+3.21}_{-3.07}$ & NC 
              & $7.58^{+4.64}_{-4.82}$ & $21.57^{+12.20}_{-14.35}$ & $1.14^{+1.29}_{-1.14}$ & NC \\
              & $-4.55^{+0.63}_{-0.67}$ & $53.18^{+2.35}_{-2.00}$ & $0.19^{+0.02}_{-0.01}$ & BC 
              & $-8.43^{+9.35}_{-16.15}$ & $35.61^{+10.50}_{-18.72}$ & $0.42^{+0.25}_{-0.42}$ & BC \\
 GM Aur & $0.50^{+1.24}_{-0.89}$ & $30.43^{+3.82}_{-5.58}$ & $3.06^{+0.77}_{-1.12}$ & NC 
        & $4.76^{+1.34}_{-1.39}$ & $48.09^{+4.69}_{-4.48}$ & $1.23^{+0.24}_{-0.23}$ & SC \\
        & $-2.89^{+47.15}_{-15.00}$ & $182.80^{+108.66}_{-92.32}$ & $0.08^{+0.10}_{-0.09}$ & BC 
        & $\compactcdots$ & $\compactcdots$ & $\compactcdots$ & $\compactcdots$ \\
 HN Tau & $-2.77^{+2.34}_{-2.34}$ & $49.78^{+1.58}_{-1.55}$ & $1.33^{+0.08}_{-0.08}$ & SC 
        & $\compactcdots$ & $\compactcdots$ & $\compactcdots$ & $\compactcdots$ \\
 MY Lup & $-4.46^{+2.40}_{-2.37}$ & $14.08^{+6.67}_{-6.32}$ & $17.16^{+16.26}_{-15.41}$ & NC 
        & $\compactcdots$ & $\compactcdots$ & $\compactcdots$ & $\compactcdots$ \\
        & $0.12^{+20.61}_{-6.47}$ & $72.91^{+27.71}_{-25.49}$ & $0.64^{+0.49}_{-0.45}$ & BC 
        & $\compactcdots$ & $\compactcdots$ & $\compactcdots$ & $\compactcdots$ \\
 RECX 11 & $-1.32^{+0.81}_{-0.91}$ & $39.77^{+2.84}_{-2.87}$ & $1.64^{+0.23}_{-0.24}$ & NC 
         & $\compactcdots$ & $\compactcdots$ & $\compactcdots$ & $\compactcdots$ \\
         & $13.61^{+8.60}_{-5.70}$ & $124.16^{+26.59}_{-16.76}$ & $0.17^{+0.07}_{-0.05}$ & BC 
         & $\compactcdots$ & $\compactcdots$ & $\compactcdots$ & $\compactcdots$ \\
 RX J1556.1 & $-5.96^{+1.41}_{-1.40}$ & $58.49^{+2.16}_{-2.15}$ & $0.31^{+0.02}_{-0.02}$ & SC 
            & $-6.43^{+2.35}_{-2.37}$ & $53.67^{+5.74}_{-5.60}$ & $0.37^{+0.08}_{-0.08}$ & SC \\
 RX J1842.9 & $-1.04^{+2.25}_{-2.26}$ & $38.42^{+2.45}_{-2.25}$ & $0.77^{+0.10}_{-0.09}$ & NC 
            & $-3.84^{+2.47}_{-2.47}$ & $43.90^{+3.19}_{-3.07}$ & $0.59^{+0.09}_{-0.08}$ & SC \\
            & $16.09^{+17.21}_{-7.71}$ & $170.23^{+55.98}_{-29.91}$ & $0.04^{+0.03}_{-0.01}$ & BC 
            & $\compactcdots$ & $\compactcdots$ & $\compactcdots$ & $\compactcdots$ \\
 RX J1852.3 & $-5.04^{+2.22}_{-2.22}$ & $14.30^{+1.16}_{-1.15}$ & $4.56^{+0.74}_{-0.73}$ & NC 
            & $-2.36^{+2.34}_{-2.35}$ & $6.62^{+4.93}_{-3.66}$ & $21.25^{+31.66}_{-21.25}$ & NC \\
            & $5.25^{+5.59}_{-4.24}$ & $84.56^{+13.30}_{-9.82}$ & $0.13^{+0.04}_{-0.03}$ & BC 
            & $5.99^{+8.25}_{-4.68}$ & $48.74^{+10.21}_{-10.08}$ & $0.39^{+0.16}_{-0.16}$ & BC \\
 RY Lup & $0.82^{+2.12}_{-2.11}$ & $43.55^{+2.15}_{-2.07}$ & $2.46^{+0.24}_{-0.23}$ & SC 
        & $\compactcdots$ & $\compactcdots$ & $\compactcdots$ & $\compactcdots$ \\
 SSTc2d J160830.7 & $-0.28^{+1.97}_{-1.97}$ & $26.47^{+1.63}_{-1.61}$ & $7.01^{+0.86}_{-0.85}$ & SC 
                  & $\compactcdots$ & $\compactcdots$ & $\compactcdots$ & $\compactcdots$ \\
 SSTc2d J161243.8 & $-2.40^{+2.53}_{-2.53}$ & $42.76^{+2.55}_{-2.41}$ & $0.42^{+0.05}_{-0.05}$ & SC 
                  & $\compactcdots$ & $\compactcdots$ & $\compactcdots$ & $\compactcdots$ \\
 SY Cha & $-22.49^{+2.06}_{-1.81}$ & $43.49^{+7.48}_{-11.98}$ & $0.88^{+0.30}_{-0.49}$ & NC 
        & $\compactcdots$ & $\compactcdots$ & $\compactcdots$ & $\compactcdots$ \\
        & $-8.72^{+13.38}_{-6.05}$ & $80.62^{+13.74}_{-9.92}$ & $0.26^{+0.09}_{-0.06}$ & BC 
        & $\compactcdots$ & $\compactcdots$ & $\compactcdots$ & $\compactcdots$ \\
 Sz 45 & $-0.68^{+0.86}_{-0.86}$ & $30.12^{+4.10}_{-5.53}$ & $1.01^{+0.28}_{-0.37}$ & NC 
       & $\compactcdots$ & $\compactcdots$ & $\compactcdots$ & $\compactcdots$ \\
       & $17.46^{+30.36}_{-10.16}$ & $83.92^{+28.68}_{-22.25}$ & $0.13^{+0.09}_{-0.07}$ & BC 
       & $\compactcdots$ & $\compactcdots$ & $\compactcdots$ & $\compactcdots$ \\
 Sz 71 & $-6.84^{+2.46}_{-2.60}$ & $6.81^{+7.37}_{-3.88}$ & $13.28^{+28.74}_{-13.28}$ & NC 
       & $\compactcdots$ & $\compactcdots$ & $\compactcdots$ & $\compactcdots$ \\
       & $-2.96^{+3.11}_{-2.88}$ & $74.69^{+5.96}_{-5.65}$ & $0.11^{+0.02}_{-0.02}$ & BC 
       & $\compactcdots$ & $\compactcdots$ & $\compactcdots$ & $\compactcdots$ \\
 Sz 75 & $6.15^{+1.58}_{-1.59}$ & $74.25^{+2.44}_{-2.36}$ & $0.43^{+0.03}_{-0.03}$ & SC 
       & $\compactcdots$ & $\compactcdots$ & $\compactcdots$ & $\compactcdots$ \\
 Sz 82 & $8.06^{+3.86}_{-10.90}$ & $26.46^{+51.56}_{-13.40}$ & $3.11^{+12.13}_{-3.11}$ & NC 
       & $\compactcdots$ & $\compactcdots$ & $\compactcdots$ & $\compactcdots$ \\
       & $-7.13^{+4.10}_{-3.60}$ & $89.06^{+5.95}_{-21.06}$ & $0.27^{+0.04}_{-0.13}$ & BC 
       & $\compactcdots$ & $\compactcdots$ & $\compactcdots$ & $\compactcdots$ \\
 Sz 98 & $-3.85^{+2.30}_{-2.29}$ & $23.10^{+5.21}_{-5.84}$ & $2.55^{+1.15}_{-1.29}$ & NC 
       & $\compactcdots$ & $\compactcdots$ & $\compactcdots$ & $\compactcdots$ \\
       & $-7.70^{+2.81}_{-3.23}$ & $82.09^{+11.43}_{-8.80}$ & $0.20^{+0.06}_{-0.04}$ & BC 
       & $\compactcdots$ & $\compactcdots$ & $\compactcdots$ & $\compactcdots$ \\
 Sz 100 & $-11.96^{+2.59}_{-2.57}$ & $18.31^{+4.25}_{-4.47}$ & $0.79^{+0.37}_{-0.39}$ & NC 
        & $\compactcdots$ & $\compactcdots$ & $\compactcdots$ & $\compactcdots$ \\
        & $-4.47^{+5.61}_{-4.42}$ & $76.83^{+15.39}_{-10.83}$ & $0.05^{+0.02}_{-0.01}$ & BC 
        & $\compactcdots$ & $\compactcdots$ & $\compactcdots$ & $\compactcdots$ \\
 Sz 102 & $-17.97^{+12.46}_{-12.46}$ & $44.87^{+3.27}_{-3.20}$ & $0.27^{+0.04}_{-0.04}$ & SC 
        & $\compactcdots$ & $\compactcdots$ & $\compactcdots$ & $\compactcdots$ \\
 Sz 103 & $-9.48^{+2.28}_{-2.29}$ & $36.50^{+1.86}_{-1.81}$ & $0.38^{+0.04}_{-0.04}$ & SC 
        & $\compactcdots$ & $\compactcdots$ & $\compactcdots$ & $\compactcdots$ \\
 Sz 111 & $-3.50^{+2.15}_{-2.15}$ & $29.00^{+1.24}_{-1.26}$ & $1.24^{+0.11}_{-0.11}$ & SC 
        & $\compactcdots$ & $\compactcdots$ & $\compactcdots$ & $\compactcdots$ \\
 \end{tabular*}
\rtask{table:7}
\end{table*}

\begin{table*}
\centering
\textbf{Table 7} \\
\text{(Continued)} \\
\smallskip
 \begin{tabular*}{\textwidth}{c @{\extracolsep{\fill}} cccccccc}
 \hline
 \hline
 & \multicolumn{3}{c}{[1,7]} & & \multicolumn{3}{c}{[3,16]} \\
 \cmidrule(lr){2-4} \cmidrule(lr){6-8}
 Name & Centroid - RV & FWHM & $R$(H\textsubscript{2}) & ID & Centroid - RV & FWHM & $R$(H\textsubscript{2}) & ID \\
 & (km s\textsuperscript{-1}) & (km s\textsuperscript{-1}) & (au) & & (km s\textsuperscript{-1}) & (km s\textsuperscript{-1}) & (au) & \\
 \hline
 Sz 114 & $-5.99^{+2.52}_{-3.66}$ & $15.87^{+21.75}_{-2.99}$ & $0.03^{+0.09}_{-0.01}$ & NC 
        & $\compactcdots$ & $\compactcdots$ & $\compactcdots$ & $\compactcdots$ \\
        & $-11.18^{+6.36}_{-2.95}$ & $52.51^{+5.43}_{-33.90}$ & ${<}0.01^{+0.00}_{-0.00}$ & BC 
        & $\compactcdots$ & $\compactcdots$ & $\compactcdots$ & $\compactcdots$ \\
 Sz 129 & $-10.23^{+2.63}_{-2.64}$ & $34.09^{+4.33}_{-5.25}$ & $0.66^{+0.17}_{-0.20}$ & NC 
        & $\compactcdots$ & $\compactcdots$ & $\compactcdots$ & $\compactcdots$ \\
        & $-1.78^{+10.30}_{-5.11}$ & $78.02^{+13.67}_{-11.58}$ & $0.13^{+0.04}_{-0.04}$ & BC 
        & $\compactcdots$ & $\compactcdots$ & $\compactcdots$ & $\compactcdots$ \\
 TW Hya & $-5.45^{+0.07}_{-0.07}$ & $13.88^{+0.32}_{-0.31}$ & $0.05^{+0.00}_{-0.00}$ & SC 
        & $3.36^{+0.11}_{-0.10}$ & $19.62^{+0.38}_{-0.37}$ & $0.03^{+0.00}_{-0.00}$ & SC \\
 UX Tau A & $5.70^{+0.47}_{-0.47}$ & $22.23^{+1.05}_{-1.06}$ & $3.21^{+0.30}_{-0.31}$ & SC 
          & $8.93^{+0.95}_{-0.95}$ & $22.05^{+2.63}_{-2.77}$ & $3.27^{+0.78}_{-0.82}$ & SC \\
 V4046 Sgr & $-0.22^{+0.26}_{-0.26}$ & $41.28^{+0.37}_{-0.36}$ & $0.62^{+0.01}_{-0.01}$ & SC 
           & $-0.70^{+0.54}_{-0.53}$ & $42.10^{+0.96}_{-0.95}$ & $0.59^{+0.03}_{-0.03}$ & SC \\
 V510 Ori & $-14.67^{+0.75}_{-0.76}$ & $30.81^{+2.55}_{-2.49}$ & $\compactcdots$ & SC 
          & $\compactcdots$ & $\compactcdots$ & $\compactcdots$ & $\compactcdots$ \\
 VZ Cha & $10.24^{+1.40}_{-1.45}$ & $21.57^{+5.04}_{-5.28}$ & $1.09^{+0.51}_{-0.53}$ & NC 
        & $25.94^{+3.03}_{-3.12}$ & $30.46^{+11.99}_{-10.60}$ & $0.55^{+0.43}_{-0.38}$ & SC \\
        & $14.07^{+9.36}_{-5.75}$ & $123.94^{+34.74}_{-20.44}$ & $0.03^{+0.02}_{-0.01}$ & BC 
        & $\compactcdots$ & $\compactcdots$ & $\compactcdots$ & $\compactcdots$ \\
 \hline
 \end{tabular*}
\begin{flushleft}
\textbf{Notes.} (a) Observed centroid velocities corrected for literature values of stellar radial velocity (RV).  Errors from the fitting routine and RVs propagated through.  (b) Per the fitting routine, the FWHM is corrected for instrumental broadening introduced by the COS LSF.  (c) The emitting radius of the H\textsubscript{2}, accounting only for errors on the FWHM values.  (d) Gaussian component type: single (SC), narrow (NC), or broad (BC).
\end{flushleft}
\end{table*}

\clearpage

\bibliography{Main}

\end{document}